\documentclass{aa}  

\usepackage{graphicx}
\usepackage{xcolor}
\usepackage{mathtools}
\usepackage{txfonts}
\newcommand{\hi}{H\,{\sc i}}
\newcommand{\ci}{C\,{\sc i}}
\newcommand{\cii}{C\,{\sc ii}}
\newcommand{\oi}{O\,{\sc i}}
\usepackage[colorlinks=true,linkcolor=blue,citecolor=blue,urlcolor=blue]{hyperref}%
\begin{document}

\title{A high-redshift calibration of the [\oi]-to-\hi\ conversion factor in star-forming galaxies}


\titlerunning{The [\oi]-to-\hi\ conversion factor in high-redshift galaxies}

\author{
Sophia~N.~Wilson\inst{1,2}, Kasper~E.~Heintz\inst{1,2}, 
Páll~Jakobsson\inst{3},
Suzanne~C.~Madden\inst{4},
Darach~Watson\inst{1,2},
Georgios~Magdis\inst{1,2,5},
Francesco~Valentino\inst{6},
Thomas~R.~Greve\inst{1,5}, \&
David~Vizgan\inst{7}
}
\institute{
Cosmic Dawn Center (DAWN), Denmark
\and
Niels Bohr Institute, University of Copenhagen, Jagtvej 128, DK-2200 Copenhagen N, Denmark\\
\email{ldr934@alumni.ku.dk and keheintz@nbi.ku.dk}
\and
Centre for Astrophysics and Cosmology, Science Institute, University of Iceland, Dunhagi 5, 107 Reykjavík, Iceland
\and
AIM, CEA, CNRS, Université Paris-Saclay, Université Paris Diderot, Sorbonne Paris Cité, 91191 Gif-sur-Yvette, France
\and
DTU-Space, Technical University of Denmark, Elektrovej 327, 2800 Kgs. Lyngby, Denmark
\and
European Southern Observatory, Karl-Schwarzschild-Str. 2, D-85748 Garching bei Munchen, Germany
\and
Department of Astronomy, University of Illinois, 1002 West Green St., Urbana, IL 61801, USA
}
\authorrunning{Wilson et al.}

   \date{Received \today}

 
  \abstract{The assembly and build-up of neutral atomic hydrogen (\hi) in galaxies is one of the most fundamental processes in galaxy formation and evolution. Studying this process directly in the early universe is hindered by the weakness of the hyperfine 21-cm \hi\ line transition, impeding direct detections and measurements of the \hi\ gas masses ($M_{\rm HI}$). Here we present a new method to infer $M_{\rm HI}$ of high-redshift galaxies using neutral, atomic oxygen as a {\em proxy}. Specifically, we derive metallicity-dependent conversion factors relating the far-infrared [\oi]-$63\mu$m and [\oi]-$145\mu$m emission line luminosities and $M_{\rm HI}$ in star-forming galaxies at $z\approx 2-6$ using gamma-ray bursts (GRBs) as probes. We substantiate these results by observations of galaxies at $z\approx 0$ with direct measurements of $M_{\rm HI}$ and [\oi]-$63\mu$m and [\oi]-$145\mu$m line luminosities in addition to hydrodynamical simulations at similar epochs. We find that the [\oi]$_{\rm 63\mu m}$-to-\hi\ and [\oi]$_{\rm 145\mu m}$-to-\hi\ conversion factors, here denoted $\beta_{\rm [OI]-63\mu m}$ and $\beta_{\rm [OI]-145\mu m}$, respectively, universally appears to be anti-correlated with the gas-phase metallicity. The high-redshift GRB measurements further predict a mean ratio of $L_{\rm [OI]-63\mu m} / L_{\rm [OI]-145\mu m}=1.55\pm 0.12$ and reveal generally less excited [\cii]. The $z \approx 0$ galaxy sample also shows systematically higher $\beta_{\rm [OI]-63\mu m}$ and $\beta_{\rm [OI]-145\mu m}$ conversion factors than the GRB sample, indicating either suppressed [\oi] emission in local galaxies likely due to their lower hydrogen densities or more extended, diffuse \hi\ gas reservoirs traced by the \hi\ 21-cm. Finally, we apply these empirical calibrations to the few high-redshift detections of [\oi]-$63\mu$m and [\oi]-$145\mu$m line transitions from the literature and further discuss the applicability of these conversion factors to probe the \hi\ gas content in the dense, star-forming ISM of galaxies at $z\gtrsim 6$, well into the epoch of reionization.}


   \maketitle
%

\section{Introduction}
 
The neutral, atomic hydrogen (\hi) content of galaxies in the early universe is one of the most fundamental ingredients in the overall process of galaxy formation and evolution \citep{Keres05,Schaye10,Dayal18}. Constraining the abundance of \hi\ in galaxies provides valuable insight into the accretion rate of neutral pristine gas from the intergalactic medium onto galaxy halos and the available gas fuel that can form molecules and subsequently stars. The \hi\ gas mass can be inferred through the hyperfine \hi\ 21-cm transition in local, massive galaxies \citep{Zwaan05,Walter08,Hoppmann15,Jones18,Catinella18}, but due to the weakness of the transition this is only feasible out to $z\lesssim 0.5$ for individual sources \citep{Fernandez16,Maddox21}. Recent efforts have pushed this out to $z\approx 1$ by measuring the combined 21-cm signal from a stack of several thousand galaxies \citep{Chowdhury20} or detected in one particular strongly lensed galaxy \citep{Chakraborty23}. However, to study the neutral gas reservoirs of more distant galaxies ($z>2$), an alternative probe of \hi\ is required.


A similar observational challenge has been encountered in the study of the {\em molecular} gas reservoirs of high-redshift galaxies \citep[see e.g.,][for reviews]{Bolatto13,Carilli13}. Molecular clouds are predominantly comprised of molecular hydrogen H$_2$, but the electronic transitions of H$_2$ are only excited at high temperatures ($T \gtrsim 10^4$\,K), much warmer than the typical temperatures observed in the cold, molecular gas phase \citep[$T\approx 20-30$\,K;][]{Weiss03,Glover16}. To circumvent this limitation, carbon monoxide (CO) or neutral-atomic carbon ([\ci]) has been used as physical tracers of H$_2$ to determine the molecular gas mass of high-redshift galaxies \citep{Tacconi10,Tacconi13,Walter11,Magdis12,Genzel15,Valentino18,Crocker19,Heintz20}. Since CO and [\ci] can only be shielded from the UV radiation field of the interstellar medium (ISM) by abundant H$_2$ molecules, these tracers are uniquely linked to molecular gas. Measuring the relative abundance of CO- or [\ci]-to-H$_2$ in these clouds thus enables estimates of the expected molecular gas mass from these tracers \citep{Bolatto13}. 

To infer the neutral {\em atomic} gas mass of high-redshift galaxies, we thus have an incentive to establish a similar proxy for \hi. Recent efforts have attempted to calibrate the far-infrared fine-structure transition of singly-ionized carbon [\cii]-$158\mu$m to the \hi\ gas mass using gamma-ray burst (GRB) sightlines through star-forming galaxies \citep{Heintz21}. While these narrow pencil-beam sightlines do not provide information about the total gas mass of the absorbing galaxies, they can accurately determine the {\em relative} abundance between [\cii] and \hi. This [\cii]-to-\hi\ calibration has subsequently been investigated with hydrodynamical simulations of galaxies at $z\sim 0-6$, which were in good agreement \citep{Vizgan22,Liang23}. The [\cii]-$158\mu$m line transition is one of the strongest ISM cooling lines and likely a robust tracer of \hi, as the ionization potential of neutral carbon (11.26 eV) is below that of neutral hydrogen (13.6 eV). Carbon is thus expected to primarily be in the singly-ionized state in the neutral ISM. Moreover, extended [\cii] emission has been observed spatially coincident with \hi\ 21-cm emission in nearby galaxies \citep{Madden93,Madden97} and has been shown observationally \citep{Pineda14,Croxall17,Cormier19,Tarantino21} and from simulations \citep{Franeck18,Olsen21, RamosPadilla21} to predominantly originate from the neutral gas phase of the ISM. The far-infrared [\cii]-$158\mu$m emission is further very bright and can be detected well into the epoch of reionization at $z\gtrsim 6$ \citep[][]{Smit18,Bouwens22,Fujimoto22,Heintz23}, enabling constraints of the \hi\ gas mass of galaxies at similar early epochs \citep{Heintz22}. However, [\cii] may also trace ionized gas in some environments as well \citep[e.g.,][]{Hollenbach99,RamosPadilla21,Wolfire22}, hindering a universal connection of this feature to \hi. 

Here we thus seek to establish a novel, more robust proxy for \hi. We consider the far-infrared transitions of neutral atomic oxygen; [\oi]$-63\mu$m and [\oi]$-145\mu$m. Since neutral oxygen has an ionization potential of 13.62 eV, almost identical to that of \hi, it is uniquely associated to the neutral gas-phase \citep{Hollenbach99}. The far-infrared transitions of [\oi] are thus potentially even more robust tracers of \hi\ than [\cii]. Further, in the dense, warm ISM the [\oi]$-63\mu$m transition will overtake [\cii] as the main ISM cooling line due to its critical density $n_{\rm crit}=5\times 10^{5}$\,cm$^{-3}$ \citep{Kaufman99,Narayanan17}. While [\oi]$-63\mu$m is typically as bright as [\cii] \citep{Cormier15}, the rest-frame wavelength is inaccessible to ALMA below redshifts $z\approx 4$. The [\oi]$-145\mu$m is more easily accessible, but is typically an order of magnitude weaker than [\cii] and [\oi]$-63\mu$m. As a consequence, [\oi] emission has still only been observed in a small sample of galaxies at $z>1$ \citep{Coppin12,Brisbin15,Wardlow17,Rybak20,Meyer22}. However, this is likely to change in the near future. 

We here derive and provide calibrations to infer the \hi\ gas mass of the dense, star-forming ISM of high-redshift star-forming galaxies using the far-infrared [\oi]$-63\mu$m and [\oi]$-145\mu$m transitions as proxies, based on empirical measurements and guided by hydrodynamic simulations. We have structured the paper as follows. 
In Sect.~\ref{sec:oitohi}, we detail the observations and analysis of a large sample of $z>2$ GRB afterglow spectra, used to derive the [\oi]-to-\hi\ conversion factor, and in Sect.~\ref{sec:results} we present the results. In Sect.~\ref{sec:disc} we compare our observations to predictions from hydrodynamical simulations and apply the calibration to a small sample of sources. In Sect.~\ref{sec:conc} we summarize and conclude on our work. 

Throughout the paper we assume the concordance $\Lambda$CDM cosmological model with $\Omega_{\rm m} = 0.315$, $\Omega_{\Lambda} = 0.685$, and $H_0 = 67.4$\,km\,s$^{-1}$\,Mpc$^{-1}$ \citep{Planck18}. We report the relative abundances of specific elements X and Y, [X/Y] = $\log (N_X/N_Y) - \log (N_X/N_Y)_\odot$, assuming the solar chemical abundances from \citet{Asplund21} following the recommendations by \citet{Lodders09}.

\begin{figure}[!t]
    \centering
    \includegraphics[width=9cm]{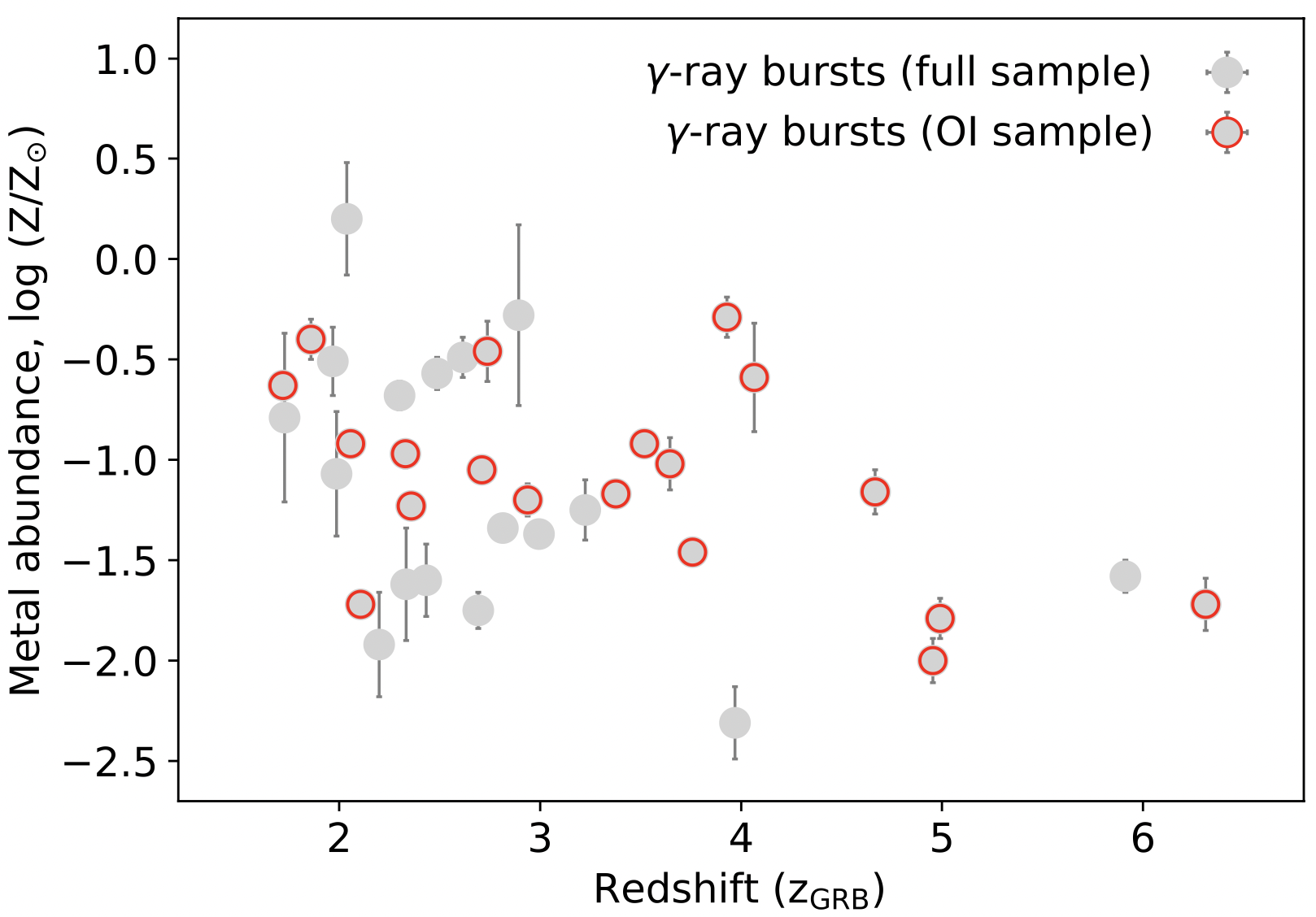}
    \caption{Gas-phase metallicities as a function of redshift. The full GRB sample from Heintz et al. (in prep.) is shown by the grey circles with the [\oi]-detected subsample highlighted by the red circles. The [\oi] sample is representative of the full underlying GRB metallicity-redshift distribution.}
    \label{fig:grbsample}
\end{figure}

\section{Deriving the [\oi]-to-\hi \ calibrations with GRBs} \label{sec:oitohi}

\subsection{GRB sample selection and observations}

Long-duration GRB afterglows have proven to be powerful probes for studying the ISM of their host galaxies \citep{Jakobsson04,Fynbo06,Prochaska07}. The progenitors of GRBs are thought to occur from the collapse and death of massive ($M\gtrsim 20-30\,M_\odot$) stars \citep{Yoon06,Woosley06,Cano17}, linking them directly to the redshift-dependent star formation rate density \citep[e.g.,][]{Robertson12}. Measuring the properties of their host galaxies thus provides a complementary view into the physical conditions in the ISM of star-forming galaxies. Since GRBs and their afterglows are some of the brightest transient events \citep{Gehrels09}, they enable studies of the ISM for galaxies even out to $z\gtrsim 6$ \citep{Hartoog15,Saccardi23}.  

\begin{table*}
\small
\centering
\caption{GRB line-of-sight metal abundances and the \hi, \oi*\,$\lambda 1304$ and \oi**\,$\lambda 1306$ column densities. The \hi\ column densities and dust-corrected gas-phase metallicities are partly from \cite{Bolmer19} and partly from Heintz et al. (in prep.). The \oi*\,$\lambda 1304$ and \oi**\,$\lambda 1306$ column densities are derived in this work. The last two columns, $\beta_{{\rm [OI]-63\mu m}}$ and $\beta_{{\rm [OI]-145\mu m}}$, denote the absorption-derived conversion factors per unit column for each GRB.}
	 \begin{tabular}{cccccccc}
	 	 \hline
	 	 \hline
	 	 GRB      & z           & log $N_{\text{HI}}$ (cm$^{-2}$) & log($Z/Z_{\odot}$) & log $N_{\text{OI}^{*}}$ (cm$^{-2}$) & log $N_{\text{OI}^{**}}$ (cm$^{-2}$) & $\log \beta_{{\rm [OI]-63\mu m}}$ ($M_{\odot} / L_{\odot}$) & $\log \beta_{{\rm [OI]-145\mu m}}$ ($M_{\odot} / L_{\odot}$) \\
	 	 \hline
	 	 090809A  & 	 2.7373 & 	 21.48 $\pm$ 0.07 & 	 -0.46 $\pm$ 0.15 & 	 15.28 $\pm$ 0.68 & 	 15.28 $\pm$ 0.68 & 	 0.26  $\pm$ 0.69  & 2.22 $\pm$ 0.50   \\
	 	 090926A  & 	 2.1069 & 	 21.58 $\pm$ 0.01 & 	 -1.72 $\pm$ 0.05 & 	 14.77 $\pm$ 0.06 & 	 14.77 $\pm$ 0.06 & 	 0.87  $\pm$ 0.06  & 2.19 $\pm$ 0.05   \\
	 	 100219A  & 	 4.6676 & 	 21.28 $\pm$ 0.02 & 	 -1.16 $\pm$ 0.11 & 	 13.71 $\pm$ 0.53 & 	 13.71 $\pm$ 0.53 & 	 1.63  $\pm$ 0.53  & 3.12 $\pm$ 0.89   \\
	 	 111008A  & 	 4.9910 & 	 22.39 $\pm$ 0.01 & 	 -1.79 $\pm$ 0.10 & 	 14.25 $\pm$ 0.78 & 	 14.25 $\pm$ 0.78 & 	 2.21  $\pm$ 0.78  & 2.88 $\pm$ 0.39   \\
	 	 120815A  & 	 2.3582 & 	 22.09 $\pm$ 0.01 & 	 -1.23 $\pm$ 0.03 & 	 15.17 $\pm$ 0.42 & 	 15.17 $\pm$ 0.42 & 	 0.98  $\pm$ 0.42  & 2.65 $\pm$ 0.12   \\
	 	 120909A  & 	 3.9290 & 	 21.82 $\pm$ 0.02 & 	 -0.29 $\pm$ 0.10 & 	 15.21 $\pm$ 0.56 & 	 15.21 $\pm$ 0.56 & 	 0.67  $\pm$ 0.56  & 2.16 $\pm$ 0.14   \\
	 	 130408A  & 	 3.7579 & 	 21.90 $\pm$ 0.01 & 	 -1.46 $\pm$ 0.05 & 	 14.38 $\pm$ 0.38 & 	 14.38 $\pm$ 0.38 & 	 1.58  $\pm$ 0.38  & 1.56 $\pm$ 0.24   \\
	 	 140311A  & 	 4.9550 & 	 22.30 $\pm$ 0.02 & 	 -2.00 $\pm$ 0.11 & 	 14.40 $\pm$ 0.55 & 	 14.40 $\pm$ 0.55 & 	 1.96  $\pm$ 0.55  & 3.59 $\pm$ 0.71   \\
	 	 150403A  & 	 2.0571 & 	 21.73 $\pm$ 0.02 & 	 -0.92 $\pm$ 0.05 & 	 15.49 $\pm$ 0.28 & 	 15.49 $\pm$ 0.28 & 	 0.30  $\pm$ 0.28  & 2.03 $\pm$ 0.13   \\
	 	 151021A  & 	 2.3297 & 	 22.14 $\pm$ 0.03 & 	 -0.97 $\pm$ 0.07 & 	 14.50 $\pm$ 0.41 & 	 14.50 $\pm$ 0.41 & 	 1.70  $\pm$ 0.41  & 2.63 $\pm$ 0.23   \\
	 	 151027B  & 	 4.0650 & 	 20.54 $\pm$ 0.07 & 	 -0.59 $\pm$ 0.27 & 	 14.99 $\pm$ 0.30 & 	 14.99 $\pm$ 0.30 & 	 -0.39  $\pm$ 0.31  & 1.63 $\pm$ 0.31   \\
	 	 160203A  & 	 3.5187 & 	 21.74 $\pm$ 0.02 & 	 -0.92 $\pm$ 0.04 & 	 14.50 $\pm$ 0.55 & 	 14.50 $\pm$ 0.55 & 	 1.30  $\pm$ 0.55  & 2.96 $\pm$ 0.58   \\
	 	 161023A  & 	 2.7100 & 	 20.95 $\pm$ 0.01 & 	 -1.05 $\pm$ 0.04 & 	 14.92 $\pm$ 0.10 & 	 14.92 $\pm$ 0.10 & 	 0.09  $\pm$ 0.10  & 1.76 $\pm$ 0.04   \\
	 	 170202A  & 	 3.6456 & 	 21.53 $\pm$ 0.04 & 	 -1.02 $\pm$ 0.13 & 	 14.43 $\pm$ 0.12 & 	 14.43 $\pm$ 0.12 & 	 1.16  $\pm$ 0.13  & 3.03 $\pm$ 0.32   \\
	 	 181020A  & 	 2.9379 & 	 22.24 $\pm$ 0.03 & 	 -1.20 $\pm$ 0.08 & 	 15.38 $\pm$ 0.27 & 	 15.38 $\pm$ 0.27 & 	 0.92  $\pm$ 0.27  & 3.18 $\pm$ 0.30   \\
	 	 190106A  & 	 1.8599 & 	 21.00 $\pm$ 0.04 & 	 -0.40 $\pm$ 0.10 & 	 15.40 $\pm$ 0.12 & 	 15.40 $\pm$ 0.12 & 	 -0.34  $\pm$ 0.13  & 2.67 $\pm$ 0.90   \\
	 	 190114A  & 	 3.3764 & 	 22.19 $\pm$ 0.05 & 	 -1.17 $\pm$ 0.06 & 	 15.32 $\pm$ 0.48 & 	 15.32 $\pm$ 0.48 & 	 0.93  $\pm$ 0.48  & 2.34 $\pm$ 0.34   \\
	 	 191011A  & 	 1.7204 & 	 21.65 $\pm$ 0.08 & 	 -0.63 $\pm$ 0.07 & 	 14.40 $\pm$ 0.39 & 	 14.40 $\pm$ 0.39 & 	 1.31  $\pm$ 0.39  & 2.86 $\pm$ 0.53   \\
	 	 210905A  & 	 6.3118 & 	 21.00 $\pm$ 0.02 & 	 -1.24 $\pm$ 0.10 & 	 14.73 $\pm$ 0.52 & 	 14.73 $\pm$ 0.52 & 	 0.33  $\pm$ 0.52  & 2.27 $\pm$ 0.38   \\
	 	 \hline
	 \end{tabular}
	 \label{tab:column_densities}
\end{table*}

The GRB sample used in this work is mainly adopted from the sample presented by \citep[][Heintz et al. in prep.]{Bolmer19,Heintz21}. The majority of these bursts are from the X-shooter GRB (XS-GRB) afterglow legacy survey \citep{Selsing19}. The observational strategy of this GRB afterglow survey is constructed such that the observed sample provides an unbiased representation of the underlying population of {\em Swift}-detected bursts. In order to obtain reliable column density estimates, we only consider spectra with a signal to noise ratio (S/N) in the wavelength region encompassing the two excited oxygen transitions, \oi*\,$\lambda 1304$ and \oi**\,$\lambda 1306$, of S/N$\gtrsim$3 per wavelength bin. Further, we require a simultaneous wavelength coverage and detection of the Lyman-$\alpha$ (Ly$\alpha$) absorption feature to determine the neutral hydrogen abundance in the line-of-sight, effectively limiting our sample to $z\gtrsim 1.7$ due to the atmospheric cutoff at lower wavelengths. We show the \oi-detected GRB sample in Fig.~\ref{fig:grbsample}, in addition to the underlying parent sample of GRBs. The final GRB afterglow sample considered in this work consists of $19$ bursts as summarized in Table~\ref{tab:column_densities}. The distribution of the redshifts and the metallicities in the GRB sample score p$=0.86$ and p=$0.99$ in a two-sided Student's t-test and the sample is therefore generally representative of the full underlying GRB sample. 

\begin{figure*}
    \centering
    \includegraphics[scale=0.6]{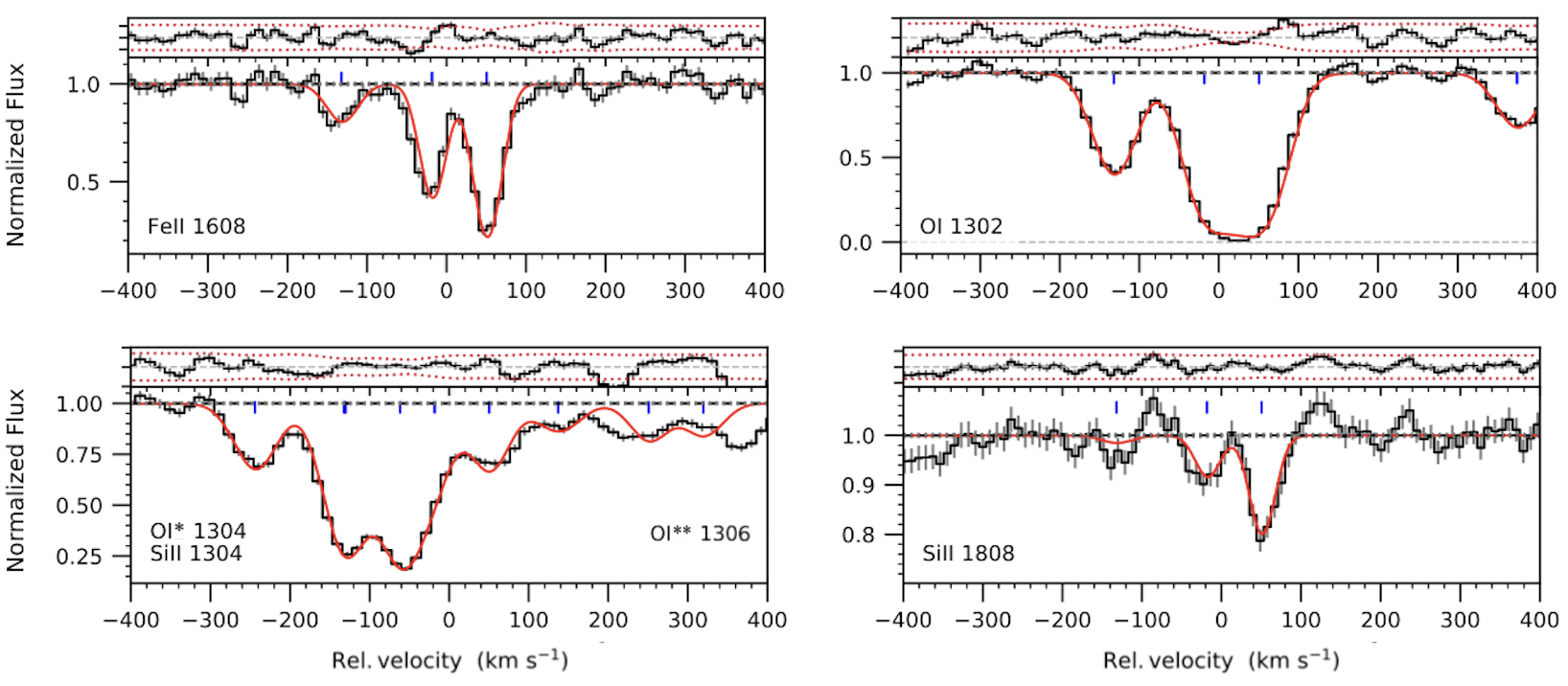}
    \caption{Representative VLT/X-shooter GRB afterglow spectrum of GRB\,161023A \citep{deUgartePostigo18}. The observed spectrum is shown in black, the solid red line represents the best-fit model of the marked absorption features. The low-ion Fe\,{\sc ii} and Si\,{\sc ii} are used to improve the modelling of the velocity components and line broadening of the \oi\,$\lambda 1302$, \oi*\,$\lambda 1304$ and \oi**\,$\lambda 1306$ line complexes}
    \label{fig:flux_velocity}
\end{figure*}

The \oi-selected bursts probe galaxies spanning redshifts from $z = 1.7204$ (GRB\,191011A) to $z = 6.3118$ (GRB\,210905A). We adopt the \hi\ column densities and dust-corrected metallicities derived by \citet[][Heintz et al. in prep]{Bolmer19} for each burst.
The gas-phase metallicities are derived as [X/Y] = $\log (N_X/N_H) - \log (N_X/N_H)_\odot$, with $(\text{X}/\text{H})_{\odot}$ representing solar abundances \citep{Asplund21}. However, since some of the metals in the interstellar medium are depleted by condensation onto interstellar dust grains, the observed depletion level [X/Y] - which is correlated with the galaxy's metallicity - is also considered \citep{DeCia16}. By correcting for the depletion level, an estimate of the total, dust-corrected metallicity $[\text{X}/\text{H}] + [\text{X}/\text{Y}] = [\text{M}/\text{H}]$ is obtained, which is represented as $\log(\text{Z}/\text{Z}_{\odot})$ (with $\log(\text{Z}/\text{Z}_{\odot})=0$ equivalent to $12+\log{\rm (O/H)}=8.69$). The GRB sample probes galaxies with \hi\ column densities in the line-of-sight in excess of $\text{N}_\text{HI}$ = $2 \times 10^{20}$\,cm$^{-2}$, classifying them as damped Lyman-$\alpha$ absorbers \citep[DLAs;][]{Wolfe05}. This ensures a large neutrality of the gas due to self-shielding, and is further representative of typical dense, star-forming regions \citep{Jakobsson06}. The relative metal abundances range from $\log (Z/Z_{\odot}) = -2.00$ (GRB\,140311A) to $\log (Z/Z_{\odot}) = -0.29$ (GRB\,120909A) (i.e. gas-phase metallicities of 1-50\% solar). 

\subsection{Absorption-line fitting} \label{ssec:abslines}

The procedure used for modelling the absorption line profiles is identical to previous work by \cite{Heintz18,Heintz21}. In this work, we determine the column densities of the excited fine-structure transitions \oi*\,$\lambda 1304$ and \oi**\,$\lambda 1306$ for each GRB in our sample. The absorption line profiles are modelled using {\tt VoigtFit} \citep{Krogager18} which fits a set of Voigt profiles to the observed absorption features and provides the redshift $z_{\rm abs}$, column density $N$ and broadening parameter $b$ as output. Both \oi* and \oi** are observed to trace the same neutral interstellar gas components as several other transitions such as Fe\,{\sc ii}, Si\,{\sc ii} and \cii\ for the GRB sightlines in our sample. This is further evidence that the excited \oi\ states predominantly traces the neutral gas-phase. We thus use these other transitions to constrain the velocity structure, number of components, and broadening parameters, when fitting the \oi*\,$\lambda 1304$ and \oi**\,$\lambda 1306$ line complexes. The intrinsic profiles are first convolved by the measured spectral resolution of each afterglow spectrum. In the case of systems where multiple velocity components are detected, representing individual gas complexes along the line-of-sight, the sum of the individual column densities is reported. This is consistent with the procedure used to measure the \hi\ abundances and the gas-phase metallicities. An example of the Voigt-profile modelling of the \oi*\,$\lambda 1304$ and \oi**\,$\lambda 1306$ transitions in GRB\,161023A is shown in Fig.~\ref{fig:flux_velocity}. The resulting column densities are listed for the full GRB \oi\ sample in Table~\ref{tab:column_densities}.

\subsection{Calibrating the [OI]-to-HI conversion factor} 

The excited fine-structure transitions \oi*\,$\lambda 1304$ and \oi**\,$\lambda 1306$ detected in absorption arise from the $^3P_{J=1}$ and $^3P_{J=0}$ levels of neutral atomic oxygen, and were explicitly chosen because they give rise to the far-infrared [\oi]$-63\mu$m ($^3P_1 \rightarrow \prescript{3}{}{P_2}$) and [\oi]$-145\mu$m ($^3P_0 \rightarrow \prescript{3}{}{P_1}$) emission line transitions, respectively. This allows us to derive the corresponding line-of-sight ``column'' luminosity of [\oi]$-63\mu$m and [\oi]$-145\mu$m from the spontaneous decay rates, expressed as $L^c_{\rm [OI]-63\mu m} = h \nu_{\text{ul}} A_{\text{ul}} N_{\rm OI^*}$ and likewise for $L^c_{\rm [OI]-145\mu m}$ for each GRB sightline \citep[see also][]{Heintz20,Heintz21}. Here, $h$ is the Planck constant, $\nu_{\text{ul}}$ and $A_{\text{ul}}$ are the line frequency and Einstein coefficient, respectively and $N$ is the column density of the excited transition. For [\oi]$-63\mu$m, $v_{\text{ul}} = 4744.8 \text{GHz}$ and $A_{\text{ul}} = 8.91 \times 10^{-5} \text{s}^{-1}$ and for [\oi]$-145\mu$m, $v_{\text{ul}} = 2060.1 \text{GHz}$ and $A_{\text{ul}} = 1.75 \times 10^{-5} \text{s}^{-1}$. Similarly, the line-of-sight \hi\ mass ``colum'' density can be determined, $M^c_{\text{HI}} = m_{\text{HI}}N_{\text{HI}}$, where $m_{\text{HI}}$ is the mass of a single hydrogen atom and $N_{\text{HI}}$ is the total \hi\ column number density. 

For each line-of-sight, we can now determine the ratios of the column line luminosities, $L^c_{\rm [OI]-63\mu m}$ and $L^c_{\rm [OI]-145\mu m}$, and relate them to the \hi\ mass column density directly as
\begin{equation}
    \beta_{\rm [OI]-63\mu m} \equiv \frac{M^c_{\text{HI}}}{L^c_{\rm [OI]-63\mu m}} = \frac{m_{\text{HI}}}{h \nu_{\text{ul}} A_{\text{ul}}} \frac{N_{\text{HI}}}{N_{\rm OI^*}}
\end{equation}
and 
\begin{equation}
    \beta_{\rm [OI]-145\mu m}  \equiv \frac{M^c_{\text{HI}}}{L^c_{\rm [OI]-145\mu m}} = \frac{m_{\text{HI}}}{h \nu_{\text{ul}} A_{\text{ul}}} \frac{N_{\text{HI}}}{N_{\rm OI^{**}}}
\end{equation}
These two expressions provide direct [\oi]$_{\rm 63\mu m}$-to-\hi\ and [\oi]$_{\rm 145\mu m}$-to-\hi\ conversion factors based on the measured column densities in the line-of-sight. Assuming that the derived ratios of the column densities for each sightline are representative of the mean of the relative total population, e.g., $N_{\text{HI}}/N_{\text{OI}^{*}} = \sum_\text{HI}/\sum_{\text{OI}^{*}}$. The calibrations derived per unit column are thus equal to the global [\oi]$_{\rm 63\mu m}$-to-\hi\ and [\oi]$_{\rm 145\mu m}$-to-\hi\ conversion factors. 

This scaling has been derived for both transitions for each GRB in the sample and converted into solar units, $M_{\odot}/L_{\odot}$, which can simply be expressed as constant factors of the absorption-derived column density ratios as
\begin{equation}
    \beta_{{\rm [OI]-63\mu m}}  = 1.150 \times 10^{-6} \frac{N_{\text{HI}}}{N_{\rm OI^*}} \frac{M_{\odot}}{L_{\odot}} 
\end{equation}
and 
\begin{equation}
    \beta_{\rm [OI]-145\mu m} = 1.138 \times 10^{-5} \frac{N_{\text{HI}}}{N_{\rm OI^{**}}}  \frac{M_{\odot}}{L_{\odot}} 
\end{equation}
We emphasize that this approach only determines the relative mass to luminosity ratios of \hi\ and the two [\oi] transitions and not the global properties for the GRB-selected galaxies. These relations thus only provide a conversion between the integrated [\oi] luminosities to the total \hi\ gas masses assuming that the GRB sightlines are representative of the overall ISM conditions. As already showcased by \citet{Heintz20} using a similar methodology to determine the [\ci]-to-H$_2$ calibration and \citet{Heintz21} for an equivalent [\cii]-to-\hi\ calibration, this methodology provides results in remarkable agreement with hydrodynamical simulations \citep{Glover16,Vizgan22,Liang23}. Further, the [\cii]-to-\hi\ scaling is also reproduced in local galaxies where the \hi\ gas mass can be constrained directly from the 21-cm hyperfine transitions \citep{RemyRuyer14,Cormier15}. This substantiates the assumption that the GRB sightline trace representative regions of the star-forming ISM. 

We derive conversion factors in the range $\log \beta_{\rm [OI]-63\mu m} =-0.389-2.205$ and $\log \beta_{\rm [OI]-145\mu m} =1.561-3.589$. The full list of measurements for each GRB sightlines is provided in Table~\ref{tab:column_densities}. As a conservative estimate, we set the systematical uncertainty for the
$\log \beta_{\rm [OI]-63\mu m}$ and $\log \beta_{\rm [OI]-145\mu m}$ measurements to 0.1\,dex and 0.2\,dex, respectively. The systematical uncertainty is higher for $\log \beta_{\rm [OI]-145\mu m}$ due to the column density of the second excited fine-structure transition often not being as well-constrained. In Sect.~\ref{sec:results} we will explore any correlations between the absorption-derived relative abundances of [\oi]$-63\mu$m and [\oi]$-145\mu$m, and their connection to the metal abundances and absorption-derived [\cii]$-158\mu $m luminosities.

\begin{figure}
    \centering
    \includegraphics[width=9cm]{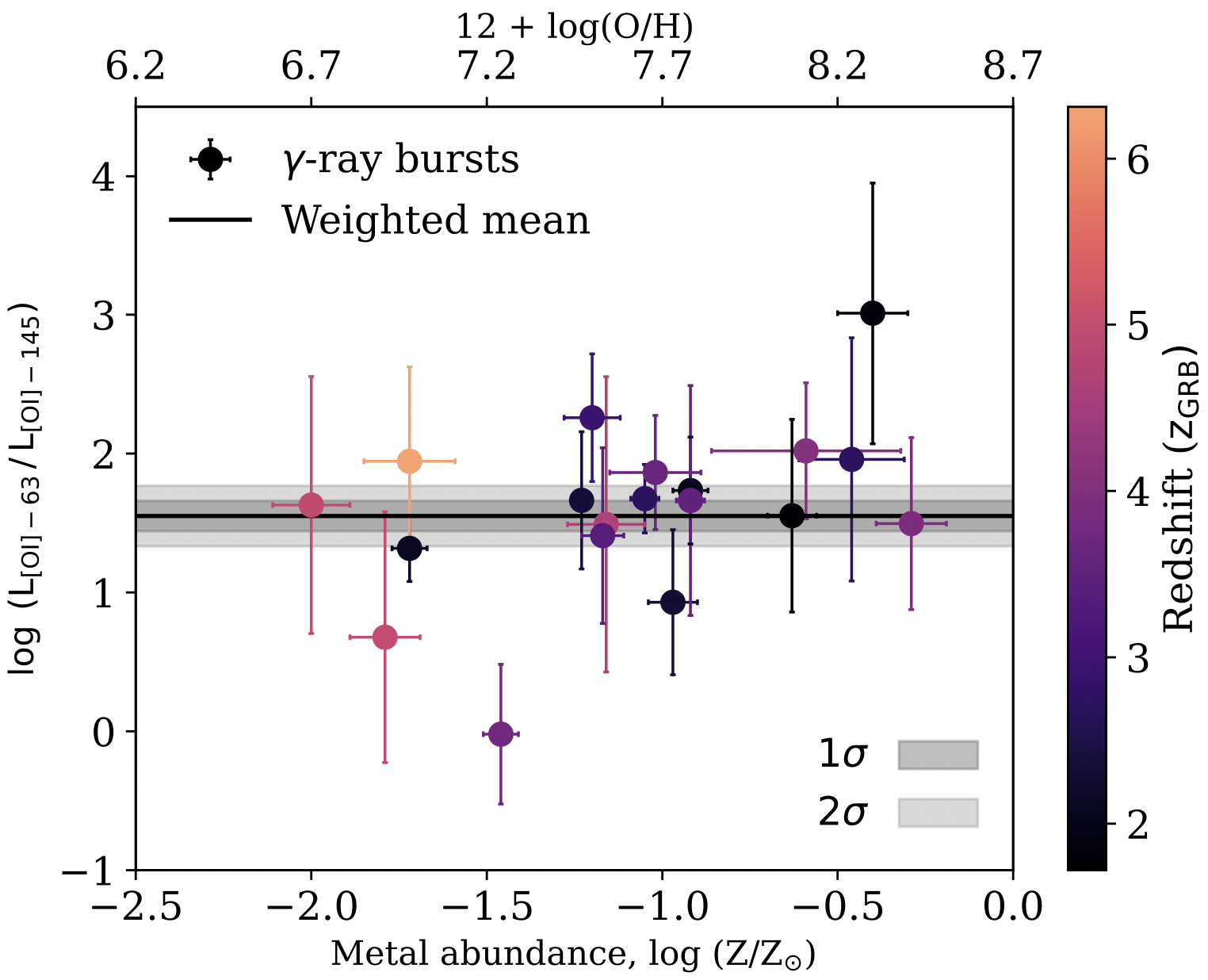}
    \caption{The luminosity ratio for the two selected transitions ([\oi]$-63\mu$m and [\oi]$-145\mu$m) in log-scale plotted against the metal abundance, $\log (Z/Z_\odot)$. The black solid line and the grey-shaded regions show the weighted mean of $\log (L_{\rm [OI]-63\mu m}/L_{\rm [OI]-145\mu m}) = 1.55 \pm 0.12$ and the associated $1$- and $2$-sigma uncertainty. The GRBs are color-coded as a function of redshift.}
    \label{fig:lumratio_met}
\end{figure}

\section{Results} \label{sec:results}


Here we first present the ratio of the absorption-derived $L_{\rm [OI]-63\mu m}$ and $L_{\rm [OI]-145\mu m}$ luminosities (i.e. the relative difference in the $\beta_{\rm [OI]-63\mu m}$ and $\beta_{\rm [OI]-145\mu m}$ calibrations) as a function of metallicity and redshift in Fig.~\ref{fig:lumratio_met}. We observe a log-linear correlation between the absorption-derived $L_{\rm [OI]-63\mu m}/L_{\rm [OI]-145\mu m}$ ratio and the gas-phase metallicity with a Pearson correlation coefficient of $\rho=0.435$ and a p-value at $p = 0.063$. There is no apparent physical explanation describing why the metal enrichment of the gas should affect the excitation states of [\oi], but it may be driven by the density and temperature of the gas \citep{Kaufman99,Hollenbach99}. The $L_{\rm [OI]-63\mu m}/L_{\rm [OI]-145\mu m}$ ratio is not observed to depend significantly on the redshift ($\rho=-0.192$), which is also indirectly related to the metallicity through the overall chemical evolution of galaxies. We thus simply compute the weighted mean, finding $\log (L_{\rm [OI]-63\mu m}/L_{\rm [OI]-145\mu m}) = 1.55 \pm 0.12$, such that [\oi]$-63\mu $m will on average be $\approx 30\times$ as bright as [\oi]$-145\mu $m, at least for galaxies with metallicities of $1-50\%$ solar. 

\begin{figure}
    \centering
    \includegraphics[width=9cm]{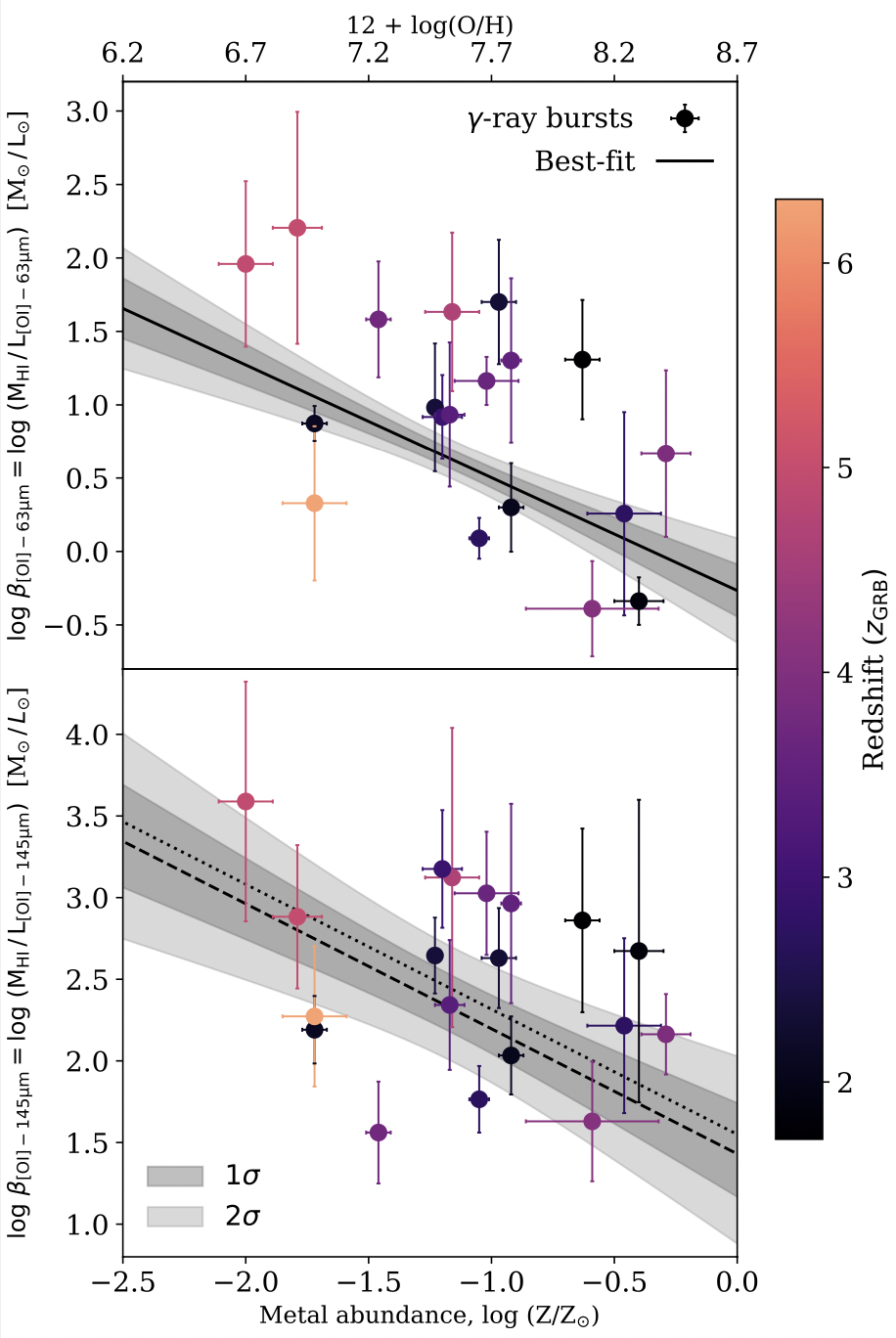}
    \caption{Absorption-derived [\oi]$_{\rm 63\mu m}$-to-\hi\ (top panel) and [\oi]$_{\rm 145\mu m}$-to-\hi\ (bottom panel) conversion factors as a function of metallicity. The color and symbol notation follow Fig.~\ref{fig:lumratio_met}. The black solid line and the grey-shaded region in the top panel represents the best fit linear relation and the associated uncertainty. The dashed line in bottom panel shows the best fit of the intersection with the slope fixed at the value of the slope from the fit in the top panel. The dotted line consist of the fit from the top panel with the weighted mean added. The grey-shaded region includes the uncertainties associated with the fit from the top panel as well as the weighted mean. }
    \label{fig:logbeta_met}
\end{figure}

\begin{figure}
    \centering
    \includegraphics[width=9cm]{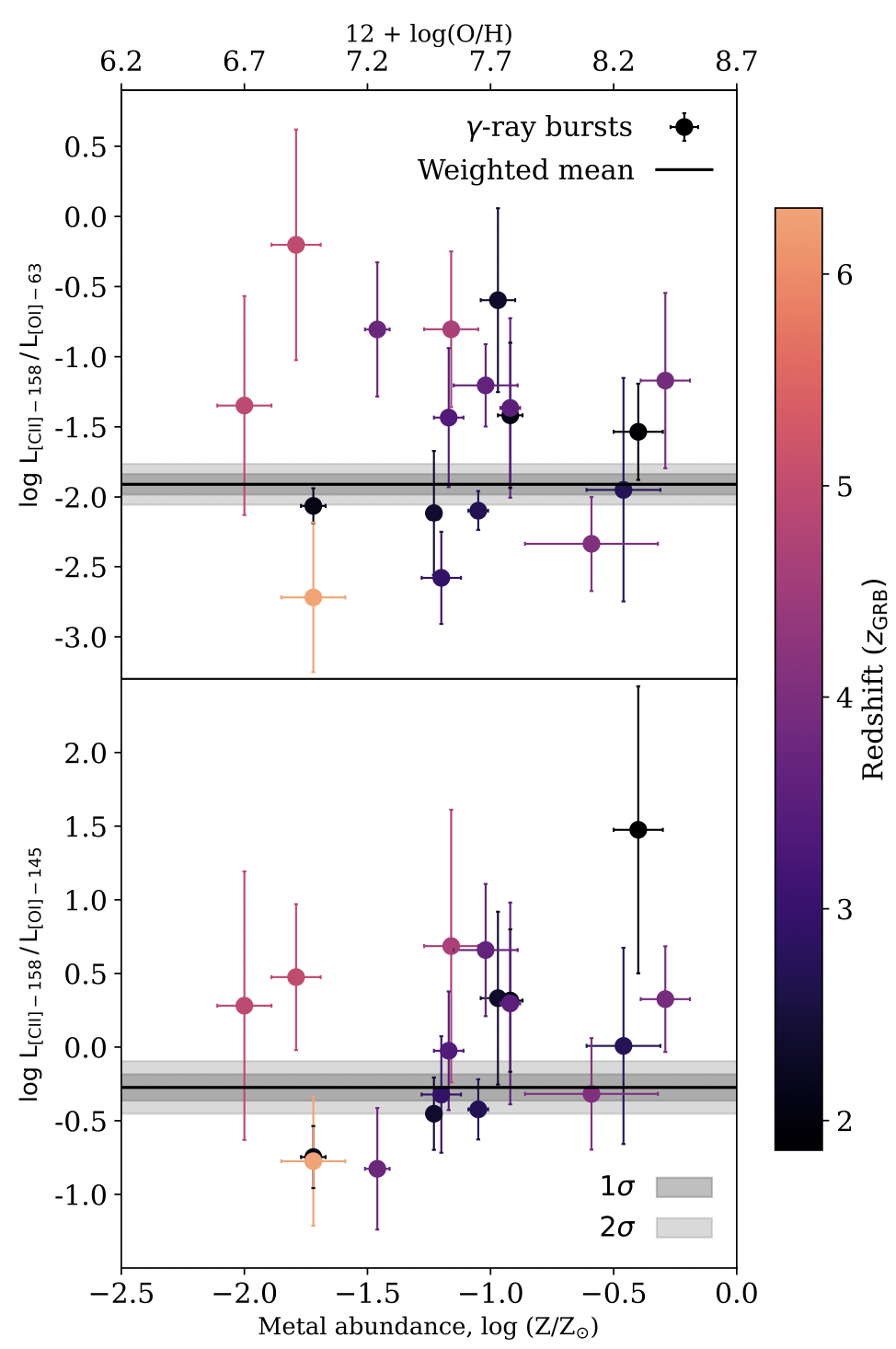}
    \caption{Absorption-derived [\oi]$_{\rm 63\mu m}$-to-[\cii]$_{\rm 158\mu m}$ and [\oi]$_{\rm 145\mu m}$-to-[\cii]$_{\rm 158\mu m}$ line luminosity ratios as a function of metallicity and redshift. The color and symbol notation follow Fig.~\ref{fig:lumratio_met}. The black solid lines and the grey-shaded regions show the weighted means and the associated $1$- and $2$-sigma uncertainty of each line ratio. }
    \label{fig:cii_oiaoib}
\end{figure}

Next, we consider the conversion factors $\log \beta_{\rm [OI]-63\mu m}$ and $\log \beta_{\rm [OI]-145\mu m}$, shown in Fig.~\ref{fig:logbeta_met} as a function of gas-phase metallicity. We observe a significant log-linear anti-correlation of $\log \beta_{\rm [OI]-63\mu m}$ with the metallicity, with a Pearson correlation coefficient of $\rho=-0.561$ and a p-value at $p=0.012$. Similarly for $\log \beta_{\rm [OI]-145\mu m}$, there appears to be a mild log-linear anti-correlation, with a Pearson correlation coefficient of $\rho=-0.258$ and a p-value at $p=0.287$. We derive best-fit [\oi]-to-\hi\ scaling relations of
\begin{equation}
\begin{multlined}
    \log (M_{\text{HI}}/M_\odot) = (-0.77 \pm 0.14) \times \log(Z/Z_{\odot}) \\ - (0.26 \pm 0.18) + \log (L_{\rm [OI]-63\mu m}/L_\odot)
\end{multlined}
\end{equation}
and
\begin{equation}
\begin{multlined}
    \log (M_{\text{HI}}/M_\odot) = (-0.77) \times \log(Z/Z_{\odot}) \\ + (1.40 \pm 0.08) + \log (L_{\rm [OI]-145\mu m}/L_\odot)~.
\end{multlined}
\end{equation}
In Eq. 6, we have fixed the slope in the [\oi]$_{\rm 145\mu m}$-to-\hi\ scaling relation to the slope found from the best-fit of [\oi]$_{\rm 63\mu m}$-to-\hi\ scaling relation (Eq.~5) and instead only fitted the intercept. This is mostly motivated by the significantly larger scatter of the $\log \beta_{\rm [OI]-145\mu m}$ measurements due to the more uncertain column density measurements of \oi**\,$\lambda 1306$, and the fact that the [\oi]$_{\rm 145\mu m}$ to [\oi]$_{\rm 63\mu m}$ ratio is theoretically expected to be constant.

\begin{figure*}
    \centering
    \includegraphics[width=1\textwidth]{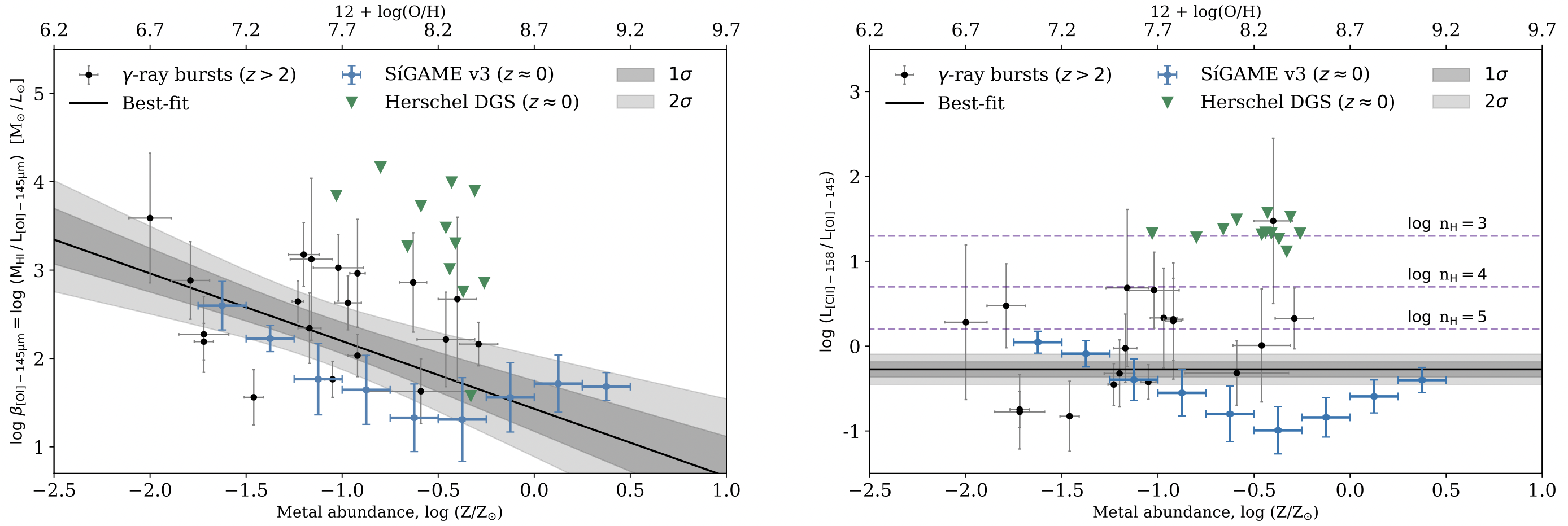}
    \caption{{\it (Left):} The [\oi]$_{\rm 145\mu m}$-to-\hi\ conversion factor as a function of metallicity. Here, the absorption-derived high-redshift GRB measurements (black dots) are compared to direct measurements from the local $z\approx 0$ {\it Herschel} Dwarf Galaxy Survey \citep{Madden13,Cormier15} and to predictions from the {\sc S\'igame} hydrodynamical simulations framework \citep{Olsen21}. Overall, the simulations are in good agreement with the GRB sightlines, whereas the low-redshift galaxies show systematically suppressed [\oi] emission per unit $M_{\rm HI}$. 
    {\it (Right):} The [\oi]$_{\rm 145\mu m}$-to-[\cii]$_{\rm 158\mu m}$ line luminosity ratio as a function of metallicity. The symbol notation is identical to the left panel. The dashed lines mark increasing volumetric hydrogen densities ($\log n_{\rm H}= 3- 5$\,cm$^{-3}$) assuming $\log U=-2$. Again, the simulations and high-redshift GRB measurements show consistent line ratios, whereas the local galaxy sample show weaker [\oi] emission for a given [\cii]-$158\mu m$ line luminosity, potentially related to their lower gas-phase densities. 
    }
    \label{fig:lumratio_cii_met}
\end{figure*}

Finally, in Fig. \ref{fig:cii_oiaoib} we show the absorption-derived [\cii]-$158\mu$m to the [\oi]-${\rm 63\mu m}$ and [\oi]-${\rm 145\mu m}$ line ratios, respectively. For this, we primarily adopt the \cii*\,$\lambda 1335.7$ measurements from \citet{Heintz21}, but also derive additional \cii*\,$\lambda 1335.7$ abundances for GRBs\,181020A, 190106A, 190114A, 191011A and 210905A following the same approach outlined in Sect.~\ref{ssec:abslines} (see Table~\ref{tab:column_densities}). We do not observe any strong correlations between the [\cii]-$158\mu$m to the [\oi]-${\rm 63\mu m}$ and [\oi]-${\rm 145\mu m}$ line ratios, with Pearson correlation coefficients of $\rho = -0.63$ and $\rho = 0.407$ and p-values at $p=0.804$ and $p=0.094$, respectively. Instead we compute weighted means of $-1.91 \pm 0.07$ and $-0.27 \pm 0.09$. These results indicate that the gas probed by the GRB sightlines generally predict weaker [\cii]-$158\mu$m emission than either of the two far-infrared [\oi] transitions. We will explore what the derived line ratios imply for the physical conditions of the gas, in addition to comparing the absorption-derived [\oi]-to-\hi\ conversion factors to complementary local galaxy samples and simulations in Sect.~\ref{sec:disc} below.


\section{Discussion} \label{sec:disc}

In an attempt to enable estimates of the \hi\ gas masses of high-redshift galaxies, we have derived the [\oi]$_{\rm 63\mu m}$-to-\hi\ and [\oi]$_{\rm 145\mu m}$-to-\hi\ conversion factors as a function of gas-phase metallicity in star-forming galaxies at $z\approx 2-6$ using GRBs as probes. To substantiate these results, we here compare the scaling relation for [\oi]-$145\mu$m in particular, to direct measurements of the $M_{\rm HI}/L_{\rm [OI]-145\mu m}$ ratios in galaxies at $z \approx 0$ in addition to predictions from recent hydrodynamical simulations. We choose the scaling relation for [\oi]-$145\mu$m since the redshiftet line is well located in the ALMA bands 6 and 7 for high-$z$ galaxies ($z>6$). In contrast, the [\oi]-$63\mu$m line is found in higher bands due to being observed at a higher frequency.



\subsection{Comparison to observations of galaxies at $z\approx 0$}

For the comparison to local galaxies at $z\approx 0$, we consider the measurements of the [\oi]-$145\mu$m line emission in addition to the \hi\ gas masses derived from the 21-cm transitions for the {\it Herschel} Dwarf Galaxy Survey (DGS) \citep{Madden13}, as presented by \citet{RemyRuyer14,Cormier15}. We also include the gas-phase metallicities reported in these works, and adopt a conservative 0.1\,dex uncertainty since these are not reported. We compare the absorption-derived [\oi]$_{\rm 145\mu m}$-to-\hi\ ratio to these local measurements as a function of metallicity in Fig.~\ref{fig:lumratio_cii_met}. The galaxies from the DGS at $z\approx 0$ appear to have systematically higher [\oi]$_{\rm 145\mu m}$-to-\hi\ abundance ratios than observed in the GRB sightlines at $z>2$. This could indicate that high-redshift galaxies potentially have less abundant \hi\ gas reservoirs per \oi\ abundance or conversely that their [\oi] emission are brighter on average, or that the local galaxy observations probe more diffuse, extended \hi\ gas via the 21-cm transition.  

Comparing also the derived [\cii]-$158\mu$m and [\oi]-${\rm 145\mu m}$ line ratios for the GRBs and the local DGS, we observe a similar suppression of [\oi] emission in the local galaxy sample in Fig.~\ref{fig:lumratio_cii_met}. Here we also show the predicted volumetric hydrogen densities, $n_{\rm H}$, for a given ionization field with $\log U = -2$ based on the $L_{\rm [CII]-158\mu m}/L_{\rm [OI]-145\mu m}$ line ratios from {\sc Cloudy} modelling of the ISM properties (Fudamoto et al. in prep.). We observe that GRBs tend to probe dense gas with $\log (n_{\rm H}/{\rm cm^{-3}}) > 4$, whereas the local galaxy sample predominantly probe more diffuse gas ($\log (n_{\rm H}/{\rm cm^{-3}}) \approx 3$).

\subsection{Predictions from hydrodynamical simulations}

In addition to the measurements from the local DGS sample, in Fig.~\ref{fig:lumratio_cii_met} we also compare our high-redshift, absorption-derived [\oi]$_{\rm 145\mu m}$-to-\hi\ conversion factor to predictions from the Simulator of Galaxy Millimeter/submillimeter Emission ({\sc S\'igame}) framework \citep[v3;][]{Olsen21}. This code provides far-infrared line emission estimates calculated through radiative transfer and physically motivated recipes from a particle-based cosmological hydrodynamics simulation, with this particular version of the code being for galaxies at $z\approx 0$.
We consider the ``neutral'' gas component, in addition to the gas-weighted metallicities from the simulation and show the average values in Fig.~\ref{fig:lumratio_cii_met} in bins of 0.25\,dex in metallicity. We find a remarkable agreement between our absorption-derived, best-fit [\oi]$_{\rm 145\mu m}$-to-\hi\ conversion factor and the predictions from {\sc S\'igame} over two orders of magnitude in gas-phase metallicity. 
Further, the observed [\cii]$-158\mu$m and [\oi]-${\rm 145\mu m}$ line ratios are also in good agreement with our results, and potentially reveal even denser gas with $\log (n_{\rm H}/{\rm cm^{-3}}) > 5$ encoded in the {\sc S\'igame} simulations. 

\subsection{Application to [\oi]-emitting galaxies at high redshifts}

The conversion factors derived here are observed either in absorption or directly in emission from the ISM of ``normal'' main-sequence star-forming galaxies, albeit over a large span in redshift and metallicity. These scaling relations are therefore only applicable to sources with similar ISM properties. However, to-date no observations of [\oi] have been published for typical star-forming galaxies at high-redshift. Instead, we attempt to infer the \hi\ gas masses of the gravitationally lensed dusty star-forming galaxy G09.83808 at $z = 6.027$ from \citet{Rybak20} and the quasar J1148+5251 at $z = 6.42$ from \citet{Meyer22}. Based on the reported intrinsic line luminosities of $L_{\rm [OI]-63\mu m}=(5.4\pm 1.3)\times 10^{9}\,L_\odot$ and $L_{\rm [OI]-145\mu m}=(1.1\pm 0.7)\times 10^{9}\,L_\odot$, we estimate $M_{\rm HI} \approx 3.0\times 10^{9}\,M_\odot$ and $M_{\rm HI} \approx 2.5\times 10^{10}\,M_\odot$, respectively, assuming solar metallicities. If the metallicities are lower (higher), these should be treated as lower (upper) limits. Intriguingly, the inferred \hi\ gas mass for the quasar J1148+5251 implies a gas-to-dust mass ratio of $\delta_{\rm DGR}\approx 100$, consistent with the Galactic average.

We emphasize though that these \hi\ gas mass estimates are still uncertain and potentially overall overestimated due to the typical lower $\alpha_{\rm CO}$ conversion factors observed in similar ULIRG type galaxies due to their stronger excitation states \citep{Papadopoulos12,Bolatto13,Accurso17}. However, in the near future we are likely to expect an increasing number of [\oi] detections from ``regular'' star-forming galaxies at $z>6$. In particular, the [\oi]-$63\mu$m transition will be redshifted into the ALMA band 9 and the [\oi]-$145\mu$m transition will be placed in band 7, enabling measurements of these important neutral gas tracers with few hours of on-target integration for the brightest [\oi]-emitting sources. To fully optimize the use of the derived [\oi]$_{\rm 63\mu m}$-to-\hi\ and [\oi]$_{\rm 145\mu m}$-to-\hi\ gas tracers, complementary {\it JWST} observations are needed to provide constraints on the metallicity of the sources at $z>6$ \citep[see e.g.,][]{Heintz23}.

\section{Conclusions} \label{sec:conc}

In this work we presented the first high-redshift calibrations of the [\oi]$_{\rm 63\mu m}$-to-\hi\ and [\oi]$_{\rm 145\mu m}$-to-\hi\ conversion factors, here denoted $\beta_{\rm [OI]-63\mu m}$ and $\beta_{\rm [OI]-145\mu m}$, respectively. This work was made in continuation of recent attempts to establish a similar [\cii]-to-\hi\ calibration \citep[e.g.,][]{Heintz21,Heintz22}. Due to the weakness of the direct 21-cm \hi\ gas tracers, such calibrations are the only alternative, and therefore vital, to infer the neutral atomic gas content of the most distant galaxies. While the far-infrared [\cii]-$158\mu$m transition is typically the brightest of the ISM cooling lines, this feature can originate from both neutral and ionized gas, making it a less optimal tracer of the neutral atomic gas only. The ionization potential and critical density of [\oi] on the other hand, ensures that the far-infrared [\oi]-$63\mu$m and [\oi]-$145\mu$m transitions predominantly originates from the neutral ISM \citep{Hollenbach99}. 


We calibrate the $\beta_{\rm [OI]-63\mu m}$ and $\beta_{\rm [OI]-145\mu m}$ conversion factors using GRBs as probes of the dense, star-forming ISM in their host galaxies, spanning redshifts $z=1.7 - 6.3$ and gas-phase metallicities from 1-50\% solar. We derive the calibrations from the measured column densities of the excited \oi*\,$\lambda 1304$ and \oi**\,$\lambda 1306$ transitions detected in absorption, which give rise to the far-infrared [\oi]-$63\mu$m and [\oi]-$145\mu$m emission lines, and from Lyman-$\alpha$ for \hi. The excited \oi\ transitions provide a measure of the luminosity per unit column assuming spontaneous decay, and Lyman-$\alpha$ the \hi\ mass per unit column. We found that both [\oi]-to-\hi\ conversion factors are anti-correlated with the gas-phase metallicity, with best-fit scaling relations:
\begin{equation*}
\begin{multlined}
    \log (M_{\text{HI}}/M_\odot) = (-0.77 \pm 0.14) \times \log(Z/Z_{\odot}) \\ - (0.26 \pm 0.18) + \log (L_{\rm [OI]-63\mu m}/L_\odot)
\end{multlined}
\label{eq:beta_63}
\end{equation*}
and
\begin{equation*}
\begin{multlined}
    \log (M_{\text{HI}}/M_\odot) = (-0.77) \times \log(Z/Z_{\odot}) \\ + (1.40 \pm 0.08) + \log (L_{\rm [OI]-145\mu m}/L_\odot)~,
\end{multlined}
\end{equation*}
(from Eqs. 5 and 6), indicating that galaxies with 10\% solar metallicities have approx. $\times 10$ higher \hi\ gas masses for a given [\oi] line luminosity compared to galaxies at solar metallicities.

We further made predictions for the line luminosity ratios of [\oi]-$63\mu$m, [\oi]-$145\mu$m, and [\cii]-$158\mu$m based on the absorption-derived conversion factors, for which we derived weighted means of $1.55\pm 0.12$ ($L_{\rm [OI]-63\mu m} / L_{\rm [OI]-145\mu m}$), $-1.91\pm 0.07$ ($L_{\rm [CII]-158\mu m} / L_{\rm [OI]-63\mu m}$), and $-0.27\pm 0.09$ ($L_{\rm [CII]-158\mu m} / L_{\rm [OI]-145\mu m}$. Overall, these results indicate that the [\oi] transitions might be the brightest far-infrared ISM cooling lines at high redshifts. We further substantiated our measurements by recent hydrodynamical simulations and local galaxy samples for which the \hi\ gas mass could be measured directly through the 21-cm line emission. The simulations were found to be in good agreement with our results. The dwarf galaxy sample at $z\approx 0$, however, showed suppressed [\oi] emission at any given \hi\ gas mass and metallicity, which we surmised could be related to the typical lower volumetric hydrogen gas densities of low-redshift galaxies, or conversely more extended, diffuse \hi\ gas components traced by the 21-cm line.      

While the next generation radio detection facilities such as the Square Kilometre Array (SKA) will significantly improve the sensitivity and the redshift range for which the \hi\ 21-cm line transition can be detected directly, these measurements will still be limited to the most massive galaxies and out to moderate distances only at $z\approx 1.7$ \citep{Blyth15}. Inferring the neutral gas content of the most distant galaxies thus necessitates the development and use of alternative gas tracers. We encourage further simulations and empirical observations to substantiate the high-redshift [\oi]-to-\hi\ calibrations derived here to establish a new window into the build-up of neutral gas in the star-forming of ISM galaxies in the early universe.

\begin{acknowledgements}
K.E.H. acknowledges support from the Carlsberg Foundation Reintegration Fellowship Grant CF21-0103.
The Cosmic Dawn Center (DAWN) is funded by the Danish National Research Foundation under grant No. 140. 
This work is partly based on observations collected at the European Organisation for Astronomical Research in the Southern Hemisphere.
\end{acknowledgements}

\bibliographystyle{aa}
\bibliography{ref.bib}

\begin{thebibliography}{77}
\expandafter\ifx\csname natexlab\endcsname\relax\def\natexlab#1{#1}\fi

\bibitem[{{Accurso} {et~al.}(2017){Accurso}, {Saintonge}, {Catinella},
  {Cortese}, {Dav{\'e}}, {Dunsheath}, {Genzel}, {Gracia-Carpio}, {Heckman},
  {Jimmy}, {Kramer}, {Li}, {Lutz}, {Schiminovich}, {Schuster}, {Sternberg},
  {Sturm}, {Tacconi}, {Tran}, \& {Wang}}]{Accurso17}
{Accurso}, G., {Saintonge}, A., {Catinella}, B., {et~al.} 2017, \mnras, 470,
  4750

\bibitem[{{Asplund} {et~al.}(2021){Asplund}, {Amarsi}, \&
  {Grevesse}}]{Asplund21}
{Asplund}, M., {Amarsi}, A.~M., \& {Grevesse}, N. 2021, \aap, 653, A141

\bibitem[{{Blyth} {et~al.}(2015){Blyth}, {van der Hulst}, {Verheijen},
  {Catinella}, {Fraternali}, {Haynes}, {Hess}, {Koribalski}, {Lagos}, {Meyer},
  {Obreschkow}, {Popping}, {Power}, {Verdes-Montenegro}, \& {Zwaan}}]{Blyth15}
{Blyth}, S., {van der Hulst}, T.~M., {Verheijen}, M.~A.~W., {et~al.} 2015, in
  Advancing Astrophysics with the Square Kilometre Array (AASKA14), 128

\bibitem[{{Bolatto} {et~al.}(2013){Bolatto}, {Wolfire}, \& {Leroy}}]{Bolatto13}
{Bolatto}, A.~D., {Wolfire}, M., \& {Leroy}, A.~K. 2013, \araa, 51, 207

\bibitem[{{Bolmer} {et~al.}(2019){Bolmer}, {Ledoux}, {Wiseman}, {De Cia},
  {Selsing}, {Schady}, {Greiner}, {Savaglio}, {Burgess}, {D'Elia}, {Fynbo},
  {Goldoni}, {Hartmann}, {Heintz}, {Jakobsson}, {Japelj}, {Kaper}, {Tanvir},
  {Vreeswijk}, \& {Zafar}}]{Bolmer19}
{Bolmer}, J., {Ledoux}, C., {Wiseman}, P., {et~al.} 2019, \aap, 623, A43

\bibitem[{{Bouwens} {et~al.}(2022){Bouwens}, {Smit}, {Schouws}, {Stefanon},
  {Bowler}, {Endsley}, {Gonzalez}, {Inami}, {Stark}, {Oesch}, {Hodge},
  {Aravena}, {da Cunha}, {Dayal}, {de Looze}, {Ferrara}, {Fudamoto},
  {Graziani}, {Li}, {Nanayakkara}, {Pallottini}, {Schneider}, {Sommovigo},
  {Topping}, {van der Werf}, {Algera}, {Barrufet}, {Hygate}, {Labb{\'e}},
  {Riechers}, \& {Witstok}}]{Bouwens22}
{Bouwens}, R.~J., {Smit}, R., {Schouws}, S., {et~al.} 2022, \apj, 931, 160

\bibitem[{{Brisbin} {et~al.}(2015){Brisbin}, {Ferkinhoff}, {Nikola},
  {Parshley}, {Stacey}, {Spoon}, {Hailey-Dunsheath}, \& {Verma}}]{Brisbin15}
{Brisbin}, D., {Ferkinhoff}, C., {Nikola}, T., {et~al.} 2015, \apj, 799, 13

\bibitem[{{Cano} {et~al.}(2017){Cano}, {Wang}, {Dai}, \& {Wu}}]{Cano17}
{Cano}, Z., {Wang}, S.-Q., {Dai}, Z.-G., \& {Wu}, X.-F. 2017, Advances in
  Astronomy, 2017, 8929054

\bibitem[{{Carilli} \& {Walter}(2013)}]{Carilli13}
{Carilli}, C.~L. \& {Walter}, F. 2013, \araa, 51, 105

\bibitem[{{Catinella} {et~al.}(2018){Catinella}, {Saintonge}, {Janowiecki},
  {Cortese}, {Dav{\'e}}, {Lemonias}, {Cooper}, {Schiminovich}, {Hummels},
  {Fabello}, {Ger{\'e}b}, {Kilborn}, \& {Wang}}]{Catinella18}
{Catinella}, B., {Saintonge}, A., {Janowiecki}, S., {et~al.} 2018, \mnras, 476,
  875

\bibitem[{{Chakraborty} \& {Roy}(2023)}]{Chakraborty23}
{Chakraborty}, A. \& {Roy}, N. 2023, \mnras, 519, 4074

\bibitem[{{Chowdhury} {et~al.}(2020){Chowdhury}, {Kanekar}, {Chengalur},
  {Sethi}, \& {Dwarakanath}}]{Chowdhury20}
{Chowdhury}, A., {Kanekar}, N., {Chengalur}, J.~N., {Sethi}, S., \&
  {Dwarakanath}, K.~S. 2020, \nat, 586, 369

\bibitem[{{Coppin} {et~al.}(2012){Coppin}, {Danielson}, {Geach}, {Hodge},
  {Swinbank}, {Wardlow}, {Bertoldi}, {Biggs}, {Brandt}, {Caselli}, {Chapman},
  {Dannerbauer}, {Dunlop}, {Greve}, {Hamann}, {Ivison}, {Karim}, {Knudsen},
  {Menten}, {Schinnerer}, {Smail}, {Spaans}, {Walter}, {Webb}, \& {van der
  Werf}}]{Coppin12}
{Coppin}, K.~E.~K., {Danielson}, A.~L.~R., {Geach}, J.~E., {et~al.} 2012,
  \mnras, 427, 520

\bibitem[{{Cormier} {et~al.}(2019){Cormier}, {Abel}, {Hony}, {Lebouteiller},
  {Madden}, {Polles}, {Galliano}, {De Looze}, {Galametz}, \&
  {Lambert-Huyghe}}]{Cormier19}
{Cormier}, D., {Abel}, N.~P., {Hony}, S., {et~al.} 2019, \aap, 626, A23

\bibitem[{{Cormier} {et~al.}(2015){Cormier}, {Madden}, {Lebouteiller}, {Abel},
  {Hony}, {Galliano}, {R{\'e}my-Ruyer}, {Bigiel}, {Baes}, {Boselli},
  {Chevance}, {Cooray}, {De Looze}, {Doublier}, {Galametz}, {Hughes},
  {Karczewski}, {Lee}, {Lu}, \& {Spinoglio}}]{Cormier15}
{Cormier}, D., {Madden}, S.~C., {Lebouteiller}, V., {et~al.} 2015, \aap, 578,
  A53

\bibitem[{{Crocker} {et~al.}(2019){Crocker}, {Pellegrini}, {Smith}, {Draine},
  {Wilson}, {Wolfire}, {Armus}, {Brinks}, {Dale}, {Groves}, {Herrera-Camus},
  {Hunt}, {Kennicutt}, {Murphy}, {Sandstrom}, {Schinnerer}, {Rigopoulou},
  {Rosolowsky}, \& {van der Werf}}]{Crocker19}
{Crocker}, A.~F., {Pellegrini}, E., {Smith}, J. D.~T., {et~al.} 2019, \apj,
  887, 105

\bibitem[{{Croxall} {et~al.}(2017){Croxall}, {Smith}, {Pellegrini}, {Groves},
  {Bolatto}, {Herrera-Camus}, {Sandstrom}, {Draine}, {Wolfire}, {Armus},
  {Boquien}, {Brandl}, {Dale}, {Galametz}, {Hunt}, {Kennicutt}, {Kreckel},
  {Rigopoulou}, {van der Werf}, \& {Wilson}}]{Croxall17}
{Croxall}, K.~V., {Smith}, J.~D., {Pellegrini}, E., {et~al.} 2017, \apj, 845,
  96

\bibitem[{{Dayal} \& {Ferrara}(2018)}]{Dayal18}
{Dayal}, P. \& {Ferrara}, A. 2018, \physrep, 780, 1

\bibitem[{{De Cia} {et~al.}(2016){De Cia}, {Ledoux}, {Mattsson}, {Petitjean},
  {Srianand}, {Gavignaud}, \& {Jenkins}}]{DeCia16}
{De Cia}, A., {Ledoux}, C., {Mattsson}, L., {et~al.} 2016, \aap, 596, A97

\bibitem[{{de Ugarte Postigo} {et~al.}(2018){de Ugarte Postigo}, {Th{\"o}ne},
  {Bolmer}, {Schulze}, {Mart{\'\i}n}, {Kann}, {D'Elia}, {Selsing},
  {Martin-Carrillo}, {Perley}, {Kim}, {Izzo}, {S{\'a}nchez-Ram{\'\i}rez},
  {Guidorzi}, {Klotz}, {Wiersema}, {Bauer}, {Bensch}, {Campana}, {Cano},
  {Covino}, {Coward}, {De Cia}, {de Gregorio-Monsalvo}, {De Pasquale}, {Fynbo},
  {Greiner}, {Gomboc}, {Hanlon}, {Hansen}, {Hartmann}, {Heintz}, {Jakobsson},
  {Kobayashi}, {Malesani}, {Martone}, {Meintjes}, {Micha{\l}owski}, {Mundell},
  {Murphy}, {Oates}, {Salmon}, {van Soelen}, {Tanvir}, {Turpin}, {Xu}, \&
  {Zafar}}]{deUgartePostigo18}
{de Ugarte Postigo}, A., {Th{\"o}ne}, C.~C., {Bolmer}, J., {et~al.} 2018, \aap,
  620, A119

\bibitem[{{Fern{\'a}ndez} {et~al.}(2016){Fern{\'a}ndez}, {Gim}, {van Gorkom},
  {Yun}, {Momjian}, {Popping}, {Chomiuk}, {Hess}, {Hunt}, {Kreckel}, {Lucero},
  {Maddox}, {Oosterloo}, {Pisano}, {Verheijen}, {Hales}, {Chung}, {Dodson},
  {Golap}, {Gross}, {Henning}, {Hibbard}, {Jaff{\'e}}, {Donovan Meyer},
  {Meyer}, {Sanchez-Barrantes}, {Schiminovich}, {Wicenec}, {Wilcots},
  {Bershady}, {Scoville}, {Strader}, {Tremou}, {Salinas}, \&
  {Ch{\'a}vez}}]{Fernandez16}
{Fern{\'a}ndez}, X., {Gim}, H.~B., {van Gorkom}, J.~H., {et~al.} 2016, \apjl,
  824, L1

\bibitem[{{Franeck} {et~al.}(2018){Franeck}, {Walch}, {Seifried}, {Clarke},
  {Ossenkopf-Okada}, {Glover}, {Klessen}, {Girichidis}, {Naab}, {W{\"u}nsch},
  {Clark}, {Pellegrini}, \& {Peters}}]{Franeck18}
{Franeck}, A., {Walch}, S., {Seifried}, D., {et~al.} 2018, \mnras, 481, 4277

\bibitem[{{Fujimoto} {et~al.}(2022){Fujimoto}, {Ouchi}, {Nakajima}, {Harikane},
  {Isobe}, {Brammer}, {Oguri}, {Gim{\'e}nez-Arteaga}, {Heintz}, {Kokorev},
  {Bauer}, {Ferrara}, {Kojima}, {Lagos}, {Laura}, {Schaerer}, {Shimasaku},
  {Hatsukade}, {Kohno}, {Sun}, {Valentino}, {Watson}, {Fudamoto}, {Inoue},
  {Gonz{\'a}lez-L{\'o}pez}, {Koekemoer}, {Knudsen}, {Lee}, {Magdis}, {Richard},
  {Strait}, {Sugahara}, {Tamura}, {Toft}, {Umehata}, \& {Walth}}]{Fujimoto22}
{Fujimoto}, S., {Ouchi}, M., {Nakajima}, K., {et~al.} 2022, arXiv e-prints,
  arXiv:2212.06863

\bibitem[{{Fynbo} {et~al.}(2006){Fynbo}, {Starling}, {Ledoux}, {Wiersema},
  {Th{\"o}ne}, {Sollerman}, {Jakobsson}, {Hjorth}, {Watson}, {Vreeswijk},
  {M{\o}ller}, {Rol}, {Gorosabel}, {N{\"a}r{\"a}nen}, {Wijers},
  {Bj{\"o}rnsson}, {Castro Cer{\'o}n}, {Curran}, {Hartmann}, {Holland},
  {Jensen}, {Levan}, {Limousin}, {Kouveliotou}, {Nelemans}, {Pedersen},
  {Priddey}, \& {Tanvir}}]{Fynbo06}
{Fynbo}, J.~P.~U., {Starling}, R.~L.~C., {Ledoux}, C., {et~al.} 2006, \aap,
  451, L47

\bibitem[{{Gehrels} {et~al.}(2009){Gehrels}, {Ramirez-Ruiz}, \&
  {Fox}}]{Gehrels09}
{Gehrels}, N., {Ramirez-Ruiz}, E., \& {Fox}, D.~B. 2009, \araa, 47, 567

\bibitem[{{Genzel} {et~al.}(2015){Genzel}, {Tacconi}, {Lutz}, {Saintonge},
  {Berta}, {Magnelli}, {Combes}, {Garc{\'\i}a-Burillo}, {Neri}, {Bolatto},
  {Contini}, {Lilly}, {Boissier}, {Boone}, {Bouch{\'e}}, {Bournaud}, {Burkert},
  {Carollo}, {Colina}, {Cooper}, {Cox}, {Feruglio}, {F{\"o}rster Schreiber},
  {Freundlich}, {Gracia-Carpio}, {Juneau}, {Kovac}, {Lippa}, {Naab}, {Salome},
  {Renzini}, {Sternberg}, {Walter}, {Weiner}, {Weiss}, \& {Wuyts}}]{Genzel15}
{Genzel}, R., {Tacconi}, L.~J., {Lutz}, D., {et~al.} 2015, \apj, 800, 20

\bibitem[{{Glover} \& {Smith}(2016)}]{Glover16}
{Glover}, S. C.~O. \& {Smith}, R.~J. 2016, \mnras, 462, 3011

\bibitem[{{Hartoog} {et~al.}(2015){Hartoog}, {Malesani}, {Fynbo}, {Goto},
  {Kr{\"u}hler}, {Vreeswijk}, {De Cia}, {Xu}, {M{\o}ller}, {Covino}, {D'Elia},
  {Flores}, {Goldoni}, {Hjorth}, {Jakobsson}, {Krogager}, {Kaper}, {Ledoux},
  {Levan}, {Milvang-Jensen}, {Sollerman}, {Sparre}, {Tagliaferri}, {Tanvir},
  {de Ugarte Postigo}, {Vergani}, {Wiersema}, {Datson}, {Salinas}, {Mikkelsen},
  \& {Aghanim}}]{Hartoog15}
{Hartoog}, O.~E., {Malesani}, D., {Fynbo}, J.~P.~U., {et~al.} 2015, \aap, 580,
  A139

\bibitem[{{Heintz} {et~al.}(2023){Heintz}, {Gim{\'e}nez-Arteaga}, {Fujimoto},
  {Brammer}, {Espada}, {Gillman}, {Gonz{\'a}lez-L{\'o}pez}, {Greve},
  {Harikane}, {Hatsukade}, {Knudsen}, {Koekemoer}, {Kohno}, {Kokorev}, {Lee},
  {Magdis}, {Nelson}, {Rizzo}, {Sanders}, {Schaerer}, {Shapley}, {Strait},
  {Toft}, {Valentino}, {Wel}, {Vijayan}, {Watson}, {Bauer}, {Christiansen}, \&
  {Wilson}}]{Heintz23}
{Heintz}, K.~E., {Gim{\'e}nez-Arteaga}, C., {Fujimoto}, S., {et~al.} 2023,
  \apjl, 944, L30

\bibitem[{{Heintz} {et~al.}(2022){Heintz}, {Oesch}, {Aravena}, {Bouwens},
  {Dayal}, {Ferrara}, {Fudamoto}, {Graziani}, {Inami}, {Sommovigo}, {Smit},
  {Stefanon}, {Topping}, {Pallottini}, \& {van der Werf}}]{Heintz22}
{Heintz}, K.~E., {Oesch}, P.~A., {Aravena}, M., {et~al.} 2022, \apjl, 934, L27

\bibitem[{{Heintz} \& {Watson}(2020)}]{Heintz20}
{Heintz}, K.~E. \& {Watson}, D. 2020, \apjl, 889, L7

\bibitem[{{Heintz} {et~al.}(2018){Heintz}, {Watson}, {Jakobsson}, {Fynbo},
  {Bolmer}, {Arabsalmani}, {Cano}, {Covino}, {D'Elia}, {Gomboc}, {Japelj},
  {Kaper}, {Krogager}, {Pugliese}, {S{\'a}nchez-Ram{\'\i}rez}, {Selsing},
  {Sparre}, {Tanvir}, {Th{\"o}ne}, {de Ugarte Postigo}, \&
  {Vergani}}]{Heintz18}
{Heintz}, K.~E., {Watson}, D., {Jakobsson}, P., {et~al.} 2018, \mnras, 479,
  3456

\bibitem[{{Heintz} {et~al.}(2021){Heintz}, {Watson}, {Oesch}, {Narayanan}, \&
  {Madden}}]{Heintz21}
{Heintz}, K.~E., {Watson}, D., {Oesch}, P.~A., {Narayanan}, D., \& {Madden},
  S.~C. 2021, \apj, 922, 147

\bibitem[{{Hollenbach} \& {Tielens}(1999)}]{Hollenbach99}
{Hollenbach}, D.~J. \& {Tielens}, A.~G.~G.~M. 1999, Reviews of Modern Physics,
  71, 173

\bibitem[{{Hoppmann} {et~al.}(2015){Hoppmann}, {Staveley-Smith}, {Freudling},
  {Zwaan}, {Minchin}, \& {Calabretta}}]{Hoppmann15}
{Hoppmann}, L., {Staveley-Smith}, L., {Freudling}, W., {et~al.} 2015, \mnras,
  452, 3726

\bibitem[{{Jakobsson} {et~al.}(2006){Jakobsson}, {Fynbo}, {Ledoux},
  {Vreeswijk}, {Kann}, {Hjorth}, {Priddey}, {Tanvir}, {Reichart}, {Gorosabel},
  {Klose}, {Watson}, {Sollerman}, {Fruchter}, {de Ugarte Postigo}, {Wiersema},
  {Bj{\"o}rnsson}, {Chapman}, {Th{\"o}ne}, {Pedersen}, \&
  {Jensen}}]{Jakobsson06}
{Jakobsson}, P., {Fynbo}, J.~P.~U., {Ledoux}, C., {et~al.} 2006, \aap, 460, L13

\bibitem[{{Jakobsson} {et~al.}(2004){Jakobsson}, {Hjorth}, {Fynbo},
  {Weidinger}, {Gorosabel}, {Ledoux}, {Watson}, {Bj{\"o}rnsson}, {Gudmundsson},
  {Wijers}, {M{\"o}ller}, {Pedersen}, {Sollerman}, {Henden}, {Jensen},
  {Gilmore}, {Kilmartin}, {Levan}, {Castro Cer{\'o}n}, {Castro-Tirado},
  {Fruchter}, {Kouveliotou}, {Masetti}, \& {Tanvir}}]{Jakobsson04}
{Jakobsson}, P., {Hjorth}, J., {Fynbo}, J.~P.~U., {et~al.} 2004, \aap, 427, 785

\bibitem[{{Jones} {et~al.}(2018){Jones}, {Haynes}, {Giovanelli}, \&
  {Moorman}}]{Jones18}
{Jones}, M.~G., {Haynes}, M.~P., {Giovanelli}, R., \& {Moorman}, C. 2018,
  \mnras, 477, 2

\bibitem[{{Kaufman} {et~al.}(1999){Kaufman}, {Wolfire}, {Hollenbach}, \&
  {Luhman}}]{Kaufman99}
{Kaufman}, M.~J., {Wolfire}, M.~G., {Hollenbach}, D.~J., \& {Luhman}, M.~L.
  1999, \apj, 527, 795

\bibitem[{{Kere{\v{s}}} {et~al.}(2005){Kere{\v{s}}}, {Katz}, {Weinberg}, \&
  {Dav{\'e}}}]{Keres05}
{Kere{\v{s}}}, D., {Katz}, N., {Weinberg}, D.~H., \& {Dav{\'e}}, R. 2005,
  \mnras, 363, 2

\bibitem[{{Krogager}(2018)}]{Krogager18}
{Krogager}, J.-K. 2018, arXiv e-prints, arXiv:1803.01187

\bibitem[{{Liang} {et~al.}(2023){Liang}, {Feldmann}, {Murray}, {Narayanan},
  {Hayward}, {Angl{\'e}s-Alc{\'a}zar}, {Bassini}, {Richings},
  {Faucher-Gigu{\`e}re}, {Chung}, {Chan}, {{\c{C}}atmabacak}, {Kere{\v{s}}}, \&
  {Hopkins}}]{Liang23}
{Liang}, L., {Feldmann}, R., {Murray}, N., {et~al.} 2023, arXiv e-prints,
  arXiv:2301.04149

\bibitem[{{Lodders} {et~al.}(2009){Lodders}, {Palme}, \& {Gail}}]{Lodders09}
{Lodders}, K., {Palme}, H., \& {Gail}, H.~P. 2009, Landolt B\&ouml;rnstein, 4B,
  712

\bibitem[{{Madden} {et~al.}(1993){Madden}, {Geis}, {Genzel}, {Herrmann},
  {Jackson}, {Poglitsch}, {Stacey}, \& {Townes}}]{Madden93}
{Madden}, S.~C., {Geis}, N., {Genzel}, R., {et~al.} 1993, \apj, 407, 579

\bibitem[{{Madden} {et~al.}(1997){Madden}, {Poglitsch}, {Geis}, {Stacey}, \&
  {Townes}}]{Madden97}
{Madden}, S.~C., {Poglitsch}, A., {Geis}, N., {Stacey}, G.~J., \& {Townes},
  C.~H. 1997, \apj, 483, 200

\bibitem[{{Madden} {et~al.}(2013){Madden}, {R{\'e}my-Ruyer}, {Galametz},
  {Cormier}, {Lebouteiller}, {Galliano}, {Hony}, {Bendo}, {Smith}, {Pohlen},
  {Roussel}, {Sauvage}, {Wu}, {Sturm}, {Poglitsch}, {Contursi}, {Doublier},
  {Baes}, {Barlow}, {Boselli}, {Boquien}, {Carlson}, {Ciesla}, {Cooray},
  {Cortese}, {de Looze}, {Irwin}, {Isaak}, {Kamenetzky}, {Karczewski}, {Lu},
  {MacHattie}, {O'Halloran}, {Parkin}, {Rangwala}, {Schirm}, {Schulz},
  {Spinoglio}, {Vaccari}, {Wilson}, \& {Wozniak}}]{Madden13}
{Madden}, S.~C., {R{\'e}my-Ruyer}, A., {Galametz}, M., {et~al.} 2013, \pasp,
  125, 600

\bibitem[{{Maddox} {et~al.}(2021){Maddox}, {Frank}, {Ponomareva}, {Jarvis},
  {Adams}, {Dav{\'e}}, {Oosterloo}, {Santos}, {Blyth}, {Glowacki},
  {Kraan-Korteweg}, {Mulaudzi}, {Namumba}, {Prandoni}, {Rajohnson}, {Spekkens},
  {Adams}, {Bowler}, {Collier}, {Heywood}, {Sekhar}, \& {Taylor}}]{Maddox21}
{Maddox}, N., {Frank}, B.~S., {Ponomareva}, A.~A., {et~al.} 2021, \aap, 646,
  A35

\bibitem[{{Magdis} {et~al.}(2012){Magdis}, {Daddi}, {B{\'e}thermin}, {Sargent},
  {Elbaz}, {Pannella}, {Dickinson}, {Dannerbauer}, {da Cunha}, {Walter},
  {Rigopoulou}, {Charmandaris}, {Hwang}, \& {Kartaltepe}}]{Magdis12}
{Magdis}, G.~E., {Daddi}, E., {B{\'e}thermin}, M., {et~al.} 2012, \apj, 760, 6

\bibitem[{{Meyer} {et~al.}(2022){Meyer}, {Walter}, {Cicone}, {Cox}, {Decarli},
  {Neri}, {Novak}, {Pensabene}, {Riechers}, \& {Weiss}}]{Meyer22}
{Meyer}, R.~A., {Walter}, F., {Cicone}, C., {et~al.} 2022, \apj, 927, 152

\bibitem[{{Narayanan} \& {Krumholz}(2017)}]{Narayanan17}
{Narayanan}, D. \& {Krumholz}, M.~R. 2017, \mnras, 467, 50

\bibitem[{{Olsen} {et~al.}(2021){Olsen}, {Burkhart}, {Mac Low}, {Tre{\ss}},
  {Greve}, {Vizgan}, {Motka}, {Borrow}, {Popping}, {Dav{\'e}}, {Smith}, \&
  {Narayanan}}]{Olsen21}
{Olsen}, K.~P., {Burkhart}, B., {Mac Low}, M.-M., {et~al.} 2021, \apj, 922, 88

\bibitem[{{Papadopoulos} {et~al.}(2012){Papadopoulos}, {van der Werf},
  {Xilouris}, {Isaak}, \& {Gao}}]{Papadopoulos12}
{Papadopoulos}, P.~P., {van der Werf}, P., {Xilouris}, E., {Isaak}, K.~G., \&
  {Gao}, Y. 2012, \apj, 751, 10

\bibitem[{{Pineda} {et~al.}(2014){Pineda}, {Langer}, \& {Goldsmith}}]{Pineda14}
{Pineda}, J.~L., {Langer}, W.~D., \& {Goldsmith}, P.~F. 2014, \aap, 570, A121

\bibitem[{{Planck Collaboration} {et~al.}(2020){Planck Collaboration},
  {Aghanim}, {Akrami}, {Ashdown}, {Aumont}, {Baccigalupi}, {Ballardini},
  {Banday}, {Barreiro}, {Bartolo}, {Basak}, {Battye}, {Benabed}, {Bernard},
  {Bersanelli}, {Bielewicz}, {Bock}, {Bond}, {Borrill}, {Bouchet}, {Boulanger},
  {Bucher}, {Burigana}, {Butler}, {Calabrese}, {Cardoso}, {Carron},
  {Challinor}, {Chiang}, {Chluba}, {Colombo}, {Combet}, {Contreras}, {Crill},
  {Cuttaia}, {de Bernardis}, {de Zotti}, {Delabrouille}, {Delouis}, {Di
  Valentino}, {Diego}, {Dor{\'e}}, {Douspis}, {Ducout}, {Dupac}, {Dusini},
  {Efstathiou}, {Elsner}, {En{\ss}lin}, {Eriksen}, {Fantaye}, {Farhang},
  {Fergusson}, {Fernandez-Cobos}, {Finelli}, {Forastieri}, {Frailis},
  {Fraisse}, {Franceschi}, {Frolov}, {Galeotta}, {Galli}, {Ganga},
  {G{\'e}nova-Santos}, {Gerbino}, {Ghosh}, {Gonz{\'a}lez-Nuevo}, {G{\'o}rski},
  {Gratton}, {Gruppuso}, {Gudmundsson}, {Hamann}, {Handley}, {Hansen},
  {Herranz}, {Hildebrandt}, {Hivon}, {Huang}, {Jaffe}, {Jones}, {Karakci},
  {Keih{\"a}nen}, {Keskitalo}, {Kiiveri}, {Kim}, {Kisner}, {Knox},
  {Krachmalnicoff}, {Kunz}, {Kurki-Suonio}, {Lagache}, {Lamarre}, {Lasenby},
  {Lattanzi}, {Lawrence}, {Le Jeune}, {Lemos}, {Lesgourgues}, {Levrier},
  {Lewis}, {Liguori}, {Lilje}, {Lilley}, {Lindholm}, {L{\'o}pez-Caniego},
  {Lubin}, {Ma}, {Mac{\'\i}as-P{\'e}rez}, {Maggio}, {Maino}, {Mandolesi},
  {Mangilli}, {Marcos-Caballero}, {Maris}, {Martin}, {Martinelli},
  {Mart{\'\i}nez-Gonz{\'a}lez}, {Matarrese}, {Mauri}, {McEwen}, {Meinhold},
  {Melchiorri}, {Mennella}, {Migliaccio}, {Millea}, {Mitra},
  {Miville-Desch{\^e}nes}, {Molinari}, {Montier}, {Morgante}, {Moss}, {Natoli},
  {N{\o}rgaard-Nielsen}, {Pagano}, {Paoletti}, {Partridge}, {Patanchon},
  {Peiris}, {Perrotta}, {Pettorino}, {Piacentini}, {Polastri}, {Polenta},
  {Puget}, {Rachen}, {Reinecke}, {Remazeilles}, {Renzi}, {Rocha}, {Rosset},
  {Roudier}, {Rubi{\~n}o-Mart{\'\i}n}, {Ruiz-Granados}, {Salvati}, {Sandri},
  {Savelainen}, {Scott}, {Shellard}, {Sirignano}, {Sirri}, {Spencer},
  {Sunyaev}, {Suur-Uski}, {Tauber}, {Tavagnacco}, {Tenti}, {Toffolatti},
  {Tomasi}, {Trombetti}, {Valenziano}, {Valiviita}, {Van Tent}, {Vibert},
  {Vielva}, {Villa}, {Vittorio}, {Wandelt}, {Wehus}, {White}, {White},
  {Zacchei}, \& {Zonca}}]{Planck18}
{Planck Collaboration}, {Aghanim}, N., {Akrami}, Y., {et~al.} 2020, \aap, 641,
  A6

\bibitem[{{Prochaska} {et~al.}(2007){Prochaska}, {Chen}, {Dessauges-Zavadsky},
  \& {Bloom}}]{Prochaska07}
{Prochaska}, J.~X., {Chen}, H.-W., {Dessauges-Zavadsky}, M., \& {Bloom}, J.~S.
  2007, \apj, 666, 267

\bibitem[{{Ramos Padilla} {et~al.}(2021){Ramos Padilla}, {Wang}, {Ploeckinger},
  {van der Tak}, \& {Trager}}]{RamosPadilla21}
{Ramos Padilla}, A.~F., {Wang}, L., {Ploeckinger}, S., {van der Tak}, F.~F.~S.,
  \& {Trager}, S.~C. 2021, \aap, 645, A133

\bibitem[{{R{\'e}my-Ruyer} {et~al.}(2014){R{\'e}my-Ruyer}, {Madden},
  {Galliano}, {Galametz}, {Takeuchi}, {Asano}, {Zhukovska}, {Lebouteiller},
  {Cormier}, {Jones}, {Bocchio}, {Baes}, {Bendo}, {Boquien}, {Boselli},
  {DeLooze}, {Doublier-Pritchard}, {Hughes}, {Karczewski}, \&
  {Spinoglio}}]{RemyRuyer14}
{R{\'e}my-Ruyer}, A., {Madden}, S.~C., {Galliano}, F., {et~al.} 2014, \aap,
  563, A31

\bibitem[{{Robertson} \& {Ellis}(2012)}]{Robertson12}
{Robertson}, B.~E. \& {Ellis}, R.~S. 2012, \apj, 744, 95

\bibitem[{{Rybak} {et~al.}(2020){Rybak}, {Zavala}, {Hodge}, {Casey}, \&
  {Werf}}]{Rybak20}
{Rybak}, M., {Zavala}, J.~A., {Hodge}, J.~A., {Casey}, C.~M., \& {Werf}, P.
  v.~d. 2020, \apjl, 889, L11

\bibitem[{{Saccardi} {et~al.}(2023){Saccardi}, {Vergani}, {De Cia}, {D'Elia},
  {Heintz}, {Izzo}, {Palmerio}, {Petitjean}, {Rossi}, {de Ugarte Postigo},
  {Christensen}, {Konstantopoulou}, {Levan}, {Malesani}, {M{\o}ller},
  {Ramburuth-Hurt}, {Salvaterra}, {Tanvir}, {Th{\"o}ne}, {Vejlgaard}, {Fynbo},
  {Kann}, {Schady}, {Watson}, {Wiersema}, {Campana}, {Covino}, {De Pasquale},
  {Fausey}, {Hartmann}, {van der Horst}, {Jakobsson}, {Palazzi}, {Pugliese},
  {Savaglio}, {Starling}, {Stratta}, \& {Zafar}}]{Saccardi23}
{Saccardi}, A., {Vergani}, S.~D., {De Cia}, A., {et~al.} 2023, \aap, 671, A84

\bibitem[{{Schaye} {et~al.}(2010){Schaye}, {Dalla Vecchia}, {Booth}, {Wiersma},
  {Theuns}, {Haas}, {Bertone}, {Duffy}, {McCarthy}, \& {van de
  Voort}}]{Schaye10}
{Schaye}, J., {Dalla Vecchia}, C., {Booth}, C.~M., {et~al.} 2010, \mnras, 402,
  1536

\bibitem[{{Selsing} {et~al.}(2019){Selsing}, {Malesani}, {Goldoni}, {Fynbo},
  {Kr{\"u}hler}, {Antonelli}, {Arabsalmani}, {Bolmer}, {Cano}, {Christensen},
  {Covino}, {D'Avanzo}, {D'Elia}, {De Cia}, {de Ugarte Postigo}, {Flores},
  {Friis}, {Gomboc}, {Greiner}, {Groot}, {Hammer}, {Hartoog}, {Heintz},
  {Hjorth}, {Jakobsson}, {Japelj}, {Kann}, {Kaper}, {Ledoux}, {Leloudas},
  {Levan}, {Maiorano}, {Melandri}, {Milvang-Jensen}, {Palazzi}, {Palmerio},
  {Perley}, {Pian}, {Piranomonte}, {Pugliese}, {S{\'a}nchez-Ram{\'\i}rez},
  {Savaglio}, {Schady}, {Schulze}, {Sollerman}, {Sparre}, {Tagliaferri},
  {Tanvir}, {Th{\"o}ne}, {Vergani}, {Vreeswijk}, {Watson}, {Wiersema},
  {Wijers}, {Xu}, \& {Zafar}}]{Selsing19}
{Selsing}, J., {Malesani}, D., {Goldoni}, P., {et~al.} 2019, \aap, 623, A92

\bibitem[{{Smit} {et~al.}(2018){Smit}, {Bouwens}, {Carniani}, {Oesch},
  {Labb{\'e}}, {Illingworth}, {van der Werf}, {Bradley}, {Gonzalez}, {Hodge},
  {Holwerda}, {Maiolino}, \& {Zheng}}]{Smit18}
{Smit}, R., {Bouwens}, R.~J., {Carniani}, S., {et~al.} 2018, \nat, 553, 178

\bibitem[{{Tacconi} {et~al.}(2010){Tacconi}, {Genzel}, {Neri}, {Cox}, {Cooper},
  {Shapiro}, {Bolatto}, {Bouch{\'e}}, {Bournaud}, {Burkert}, {Combes},
  {Comerford}, {Davis}, {F{\"o}rster Schreiber}, {Garcia-Burillo},
  {Gracia-Carpio}, {Lutz}, {Naab}, {Omont}, {Shapley}, {Sternberg}, \&
  {Weiner}}]{Tacconi10}
{Tacconi}, L.~J., {Genzel}, R., {Neri}, R., {et~al.} 2010, \nat, 463, 781

\bibitem[{{Tacconi} {et~al.}(2013){Tacconi}, {Neri}, {Genzel}, {Combes},
  {Bolatto}, {Cooper}, {Wuyts}, {Bournaud}, {Burkert}, {Comerford}, {Cox},
  {Davis}, {F{\"o}rster Schreiber}, {Garc{\'\i}a-Burillo}, {Gracia-Carpio},
  {Lutz}, {Naab}, {Newman}, {Omont}, {Saintonge}, {Shapiro Griffin}, {Shapley},
  {Sternberg}, \& {Weiner}}]{Tacconi13}
{Tacconi}, L.~J., {Neri}, R., {Genzel}, R., {et~al.} 2013, \apj, 768, 74

\bibitem[{{Tarantino} {et~al.}(2021){Tarantino}, {Bolatto}, {Herrera-Camus},
  {Harris}, {Wolfire}, {Buchbender}, {Croxall}, {Dale}, {Groves}, {Levy},
  {Riquelme}, {Smith}, \& {Stutzki}}]{Tarantino21}
{Tarantino}, E., {Bolatto}, A.~D., {Herrera-Camus}, R., {et~al.} 2021, \apj,
  915, 92

\bibitem[{{Valentino} {et~al.}(2018){Valentino}, {Magdis}, {Daddi}, {Liu},
  {Aravena}, {Bournaud}, {Cibinel}, {Cormier}, {Dickinson}, {Gao}, {Jin},
  {Juneau}, {Kartaltepe}, {Lee}, {Madden}, {Puglisi}, {Sanders}, \&
  {Silverman}}]{Valentino18}
{Valentino}, F., {Magdis}, G.~E., {Daddi}, E., {et~al.} 2018, \apj, 869, 27

\bibitem[{{Vizgan} {et~al.}(2022){Vizgan}, {Heintz}, {Greve}, {Narayanan},
  {Dav{\'e}}, {Olsen}, {Popping}, \& {Watson}}]{Vizgan22}
{Vizgan}, D., {Heintz}, K.~E., {Greve}, T.~R., {et~al.} 2022, \apjl, 939, L1

\bibitem[{{Walter} {et~al.}(2008){Walter}, {Brinks}, {de Blok}, {Bigiel},
  {Kennicutt}, {Thornley}, \& {Leroy}}]{Walter08}
{Walter}, F., {Brinks}, E., {de Blok}, W.~J.~G., {et~al.} 2008, \aj, 136, 2563

\bibitem[{{Walter} {et~al.}(2011){Walter}, {Wei{\ss}}, {Downes}, {Decarli}, \&
  {Henkel}}]{Walter11}
{Walter}, F., {Wei{\ss}}, A., {Downes}, D., {Decarli}, R., \& {Henkel}, C.
  2011, \apj, 730, 18

\bibitem[{{Wardlow} {et~al.}(2017){Wardlow}, {Cooray}, {Osage}, {Bourne},
  {Clements}, {Dannerbauer}, {Dunne}, {Dye}, {Eales}, {Farrah}, {Furlanetto},
  {Ibar}, {Ivison}, {Maddox}, {Micha{\l}owski}, {Riechers}, {Rigopoulou},
  {Scott}, {Smith}, {Wang}, {van der Werf}, {Valiante}, {Valtchanov}, \&
  {Verma}}]{Wardlow17}
{Wardlow}, J.~L., {Cooray}, A., {Osage}, W., {et~al.} 2017, \apj, 837, 12

\bibitem[{{Wei{\ss}} {et~al.}(2003){Wei{\ss}}, {Henkel}, {Downes}, \&
  {Walter}}]{Weiss03}
{Wei{\ss}}, A., {Henkel}, C., {Downes}, D., \& {Walter}, F. 2003, \aap, 409,
  L41

\bibitem[{{Wolfe} {et~al.}(2005){Wolfe}, {Gawiser}, \& {Prochaska}}]{Wolfe05}
{Wolfe}, A.~M., {Gawiser}, E., \& {Prochaska}, J.~X. 2005, \araa, 43, 861

\bibitem[{{Wolfire} {et~al.}(2022){Wolfire}, {Vallini}, \&
  {Chevance}}]{Wolfire22}
{Wolfire}, M.~G., {Vallini}, L., \& {Chevance}, M. 2022, \araa, 60, 247

\bibitem[{{Woosley} \& {Bloom}(2006)}]{Woosley06}
{Woosley}, S.~E. \& {Bloom}, J.~S. 2006, \araa, 44, 507

\bibitem[{{Yoon} {et~al.}(2006){Yoon}, {Langer}, \& {Norman}}]{Yoon06}
{Yoon}, S.~C., {Langer}, N., \& {Norman}, C. 2006, \aap, 460, 199

\bibitem[{{Zwaan} {et~al.}(2005){Zwaan}, {Meyer}, {Staveley-Smith}, \&
  {Webster}}]{Zwaan05}
{Zwaan}, M.~A., {Meyer}, M.~J., {Staveley-Smith}, L., \& {Webster}, R.~L. 2005,
  \mnras, 359, L30

\end{thebibliography}


\end{document}